%% file: UltraMat_Main_revision_Apr18.tex
\documentclass[12pt]{article}
\usepackage{amsfonts,amsmath,amssymb,amsthm,bbm,bm}
\usepackage[title]{appendix}
\usepackage{graphicx,psfrag,epsf,enumerate,hyperref}
\usepackage{natbib}
\usepackage[utf8]{inputenc}
\usepackage[english]{babel}
\usepackage[margin=1.7in]{geometry}
\usepackage{subcaption,lscape,url,soul,textalpha, pifont,ulem}
\usepackage[font=small,labelfont=bf]{caption}
\usepackage{float,multirow,booktabs,tabularx,makecell,comment}
\usepackage[table,xcdraw]{xcolor}
\usepackage{rotating,titlesec,enumitem}
\usepackage[noend]{algpseudocode}
\usepackage[plain,noend]{algorithm2e}
\makeatletter
\renewcommand{\algocf@captiontext}[2]{#1\algocf@typo. \AlCapFnt{}#2} 
\def\@algocf@capt@plain{top}
\renewcommand{\algocf@makecaption}[2]{%
  \addtolength{\hsize}{\algomargin}%
  \sbox\@tempboxa{\algocf@captiontext{#1}{#2}}%
  \ifdim\wd\@tempboxa >\hsize
    \hskip .5\algomargin%
    \parbox[t]{\hsize}{\algocf@captiontext{#1}{#2}}
  \else%
    \global\@minipagefalse%
    \hbox to\hsize{\box\@tempboxa}
  \fi%
  \addtolength{\hsize}{-\algomargin}%
}
\makeatother

\newcommand{\blind}{1}

\newcommand{\transp}{{\sf T}}

\theoremstyle{plain}
\newtheorem{theorem}{Theorem}
\newtheorem{lemma}{Lemma}
\newtheorem{proposition}{Proposition}
\newtheorem{corollary}{Corollary}
\theoremstyle{definition}
\newtheorem{definition}{Definition}

\newcommand{\X}{\mathcal{X}}

\usepackage{NotationCommands}

\newcommand{\footremember}[2]{%
	\footnote{#2}
	\newcounter{#1}
	\setcounter{#1}{\value{footnote}}%
}
\newcommand{\footrecall}[1]{%
	\footnotemark[\value{#1}]%
}
\begin{document}
	\setstcolor{red}

	\def\spacingset#1{\renewcommand{\baselinestretch}%
		{#1}\small\normalsize} \spacingset{1}
	\input{definitions}

 \title{Geometry-driven Bayesian Inference for Ultrametric Covariance Matrices}
		\author{
		Tsung-Hung Yao\footremember{mda}{Department of Biostatistics, The University of Texas MD Anderson Cancer Center, USA} \and
		Zhenke Wu\footremember{mich}{Department of Biostatistics, University of Michigan, USA} \and
		Karthik Bharath\footremember{nott}{School of Mathematical Sciences, University of Nottingham, UK}\and
		Veerabhadran Baladandayuthapani\footrecall{mich}
	}
\date{}
		\maketitle

        \begin{abstract}
 Ultrametric matrices are a class of covariance matrices that arise in latent tree models. As a parameter space in a statistical model, the set of ultrametric matrices is neither convex nor a smooth manifold. Focus in the literature has hitherto been restricted to estimation through projections and relaxation-based techniques, and inferential methods are lacking. Motivated by this, we establish a bijection between the set of positive definite ultrametric matrices and the set of rooted, leaf-labeled trees equipped with the stratified geometry of the well-known phylogenetic treespace. Using the pullback geometry under the bijection and by adapting sampling algorithms in Bayesian phylogenetics, we develop algorithms to sample from the posterior distribution on the set of ultrametric matrices in a Bayesian latent tree model where the tree may be binary or multifurcating. We demonstrate the utility of the algorithms in simulation studies, and illustrate them on a pre-clinical cancer application to quantify uncertainty about treatment trees that identify treatments with high mechanism similarity that target correlated pathways.

		\end{abstract}

	\spacingset{1.3} 
	
\section{Introduction}
\subsection{Background}
Structured covariance and inverse covariance matrices abound in statistical models for data exhibiting specific forms of dependencies \citep{GraphicalModels}. The focus of this work is on Gaussian models on trees parameterized by a class of structure covariance matrices known as ultrametric matrices consisting of non-negative entries. The class constitutes a special case of covariance graph models first studied by \citet{CW1} with additional constraints on the covariance matrix, and are intimately related to multivariate totally positive of order two distributions \citep{MTP2_InvM_KARLIN1983, MLE_MTP2}. Gaussian models parameterized by ultrametric matrices are also referred to as Gaussian tree models \citep{JF} and latent tree models \citep{WZC, CTAW}, and are widely employed in various application domains, including 
psychology \citep{MLE_MTP2}, cancer biology \citep{yao2022probabilistic}, and finance \citep{10.1093/jjfinec/nbaa018}; see \citet{PZ} for a comprehensive survey of latent tree models. 

Ultrametric matrices first appeared in the potential theory of finite state Markov chains \citep{InvMMat_UltraMat}. They have since appeared in the literature as tree-structured matrices \citep{McCullagh2006StructuredCM, MIP_Tr}, and their inverses have been used 
in network analysis \citep{Fang2023}, and in analyzing the microbial tree of life \citep{gorman2023sparsification}. Furthermore, the class of the ultrametric matrices arises naturally as the pairwise distance matrix for the elements in the ultrametric space. An important ultrametric space is the space of p-adic number, which is heavily used in modern physics, biology, cryptography, etc \citep{pAdic_Rozikov2013, pAdic_Murtagh2012, pAdic_Dragovich2017}.
Their importance in data analysis is characterized by the observation that the matrix of normalized Euclidean distances between points in some random subsets of $\mathbb R^d$ converges in probability to an ultrametric matrix as $d \to \infty$ \citep{zubarev2014stochastic,zubarev2017ultrametric}.

The set $\tilde \cU_p$ of $p \times p$ ultrametric matrices consists of both singular and nonsingular matrices. It is characterized by semialgebraic inequalities, and, aided by techniques from combinatorics and polyhedral geometry, there is a substantial body of work that focuses on their estimation. Work on developing inferential methods for them, however, has not experienced the same level of progress. Central to the set of challenges that are stymieing progress for inference is the non-trivial geometry of $ \tilde \cU_p$. This can be intuited through the link between an ultrametric matrix $\Sigma^T \in \tilde \cU_p$ and a rooted tree $T$ with $p$ leaves: the tree $T$ possesses both topological (graph structure) and geometric (edge lengths) information that is encoded in $\Sigma^T$, and varying $T$ by altering both sources of information varies the structure of $\Sigma^T$ in a non-trivial manner. 

\subsection{Related work, and contributions}
Ultrametric matrices and their generalizations \citep[e.g][]{HoAne2014, Cavicchia2022} have been studied via the corresponding set of metric trees $\cT$ they encode, not necessarily bijective with $\tilde \cU_p$. On the other hand, bijections between a class of binary trees on $p$ leaves and $\tilde \cU_p$ have been determined \citep[e.g.][Chapter 7]{semple2003phylogenetics}, but an explicit geometry that facilitates sampling and optimization is unavailable. 

Embedded geometries within symmetric matrices for a class of correlation matrices related to ultrametric matrices, with entries in $[0,1]$, via their link to tree-based metrics on $\{1,2,\ldots,p\}$ \citep{PB}, have been developed \citep{JK, MS, GNLH,LGNH}. \citet{SZAS} used an embedded geometry to assess adequacy of a Gaussian tree model for data, however with an inverse Wishart prior with support on all positive definite matrices, not necessarily those that encode tree structures. Based on their work, \citet{LD} developed a bootstrap-based procedure for testing for a more general class of latent tree models. We are unaware of such embeddings for $\tilde \cU_p$. 

Broadly, the state-of-the-art estimation methods on $\tilde \cU_p$ are two types: (i) assume that topology of the encoded tree $T$ within $\Sigma^T$ is known, and thus treat $\Sigma^T$ as a linear covariance matrix \citep{TWA} expressed in a fixed \emph{known} basis \citep{ZUR, Sturmfels2021}; (ii) project a pilot estimator of $\Sigma^T$ onto $\tilde \cU_p$ \citep{McCullagh2006StructuredCM, MIP_Tr}. The first type is restrictive, while the second suffers from not accounting for the intrinsic geometry of $\tilde \cU_p$ since it is not convex. While asymptotic inference is in principle possible for type (i), there are presently no methods for inference for type (ii). 

Within this context, our main contribution lies in establishing that the subset $\cU_p$ of positive definite ultrametric matrices within $\tilde \cU_p$ can be mapped bijectively to the well-studied phylogenetic treespace \citep{Billera2001}, and in demonstrating how this bijection may be used to pullback the CAT(0) stratified geometry of the treespace and construct efficient Markov Chain Monte Carlo algorithms on $\cU_p$ by leveraging existing methods from the phylogenetics literature. Specifically, we obtain a novel representation of an ultrametric matrix $\Sigma^T$ as a linear combination of a set of binary basis matrices, which characterizes the set of possible topologies for the tree $T$ and elucidates on how topology of the tree $T$ is encoded via the choice of the stratum and edge lengths determine its location within a stratum; a stratified geometry for $\cU_p$ was first noted by \citet{McCullagh2006StructuredCM} without details. The uncovered bijection thus enable us to simultaneously handle $\Sigma^T$ that encode binary and multifurcating trees $T$.

Since a tree is also a graph, we note that the class of Gaussian models parameterized by covariance matrices that are ultrametric is subsumed under the class of Gaussian covariance graph models, where (in)dependence between nodes are specified through zeros of the covariance matrix and not its inverse. General conjugate and non-conjugate Bayesian methods developed \citep{DL, LM, KR} for such models based on prior distributions on structured covariance matrices with pre-specified zeros can in principle be adapted to the covariances in $\mathcal U_p$. However, information on how topology and geometry of the tree $T$ manifests in  $\Sigma^T$ is not transparent. 

The rest of the paper is organized as follows. Section \ref{sec:ultrametric} reviews ultrametric matrices and properties of $\mathcal U_p$, and section \ref{sec:BHV} reviews the phylogenetic treespace of \cite{Billera2001}. Section \ref{sec:geometry} establishes the bijection between $\cU_p$ and the treespace, describes an intrinsic geometry of $\mathcal U_p$ (Theorem \ref{thm1}) and the ensuing representation of an ultrametric matrix (Corollary \ref{cor1}), and discusses how the stratified geometry of $\mathcal U_p$ with identified boundaries manifests in the ultrametric matrices (Proposition \ref{prop1}). Section \ref{sec:compute_bijection} presents an algorithm to compute the bijection. In Section \ref{sec:TrPrior}, we construct a Markovian and consistent prior on $\cU_p$. Section \ref{sec:Inference} describes an algorithm (Algorithm \ref{algo_MH}) that draws posterior covariances effectively via the intrinsic geometry of $\mathcal U_p$. In Section \ref{sec:Simulation}, we present results from extensive simulations that assess quality of recovery of an ultrametric matrix covariance in Gaussian and misspecified models, along with uncertainty quantification. In Section \ref{sec:DataAnalysis}, we demonstrate utility of the developed  Bayesian algorithm on pre-clinical data obtained from studies on cancer treatments. Section \ref{sec:Discussion} offers concluding remarks and future directions. General purpose code in \texttt{R} with packages and datasets for the proposed method is available at \url{https://github.com/bayesrx/ultrametricMat}. 

\section{Preliminaries}
\label{sec:TrCov}
In this Section, we first introduce and review properties of ultrametric matrices, and the challenges involved in their estimation and inference, and then briefly review the phylogenetic treespace by \cite{Billera2001}, known as the BHV space, and its salient properties.  

\subsection{Ultrametric matrices}
\label{sec:ultrametric}
Define $[p]:=\{1,\ldots,p\}$. Let $\mathcal S_p$ be the set of $p \times p$ real symmetric matrices and $L:=\{0,1,2,\ldots,p\}$.
\begin{definition}
A matrix $\Sigma^T \in \mathcal S_p$ is strictly ultrametric if
\begin{enumerate}
    \item [(i)] every element of $\Sigma^T$ is non-negative and diagonal elements are positive; 
    \item [(ii)] for all $i \in [p]$, $\Sigma^T_{ii}>\max\{\Sigma^T_{ij}: j\in [p]\setminus\{i\}\}$;
    \item [(iii)] for all $i, j, k \in [p]$
    \begin{equation}
    \label{UltraIneq}
        \Sigma^T_{ij} \geq \min\{\Sigma^T_{ik} ,\Sigma^T_{kj} \}\thickspace.
    \end{equation}
\end{enumerate}
\end{definition}
If property (ii) in the definition above is relaxed to ``$\geq$", then $\Sigma^T$ is referred to as an ultrametric matrix \citep{InvMMat_UltraMat}. Matrices that satisfy even weaker versions of (i)--(iii) have also been considered in the literature; see, for example, matrices with a `3-point structure' \citep{HoAne2014} and extended ultrametric matrices \citep{cavicchia2022gaussian}. Strict inequality in (ii) ensures that every strict ultrametric matrix is nonsingular and positive definite, since it is the inverse of a nonsingular matrix with non-positive off-diagonal entries (known as an M-matrix) \citep[e.g.][Theorem 3.5]{InvMMat_UltraMat}. Motivated by the use of nonsingular covariance matrices in latent tree models, and by the application in Section \ref{sec:DataAnalysis}, we focus on strictly ultrametric matrices.  Denote by $\mathcal U_p$ the set of $p \times p$ strict ultrametric matrices, a subset of the convex cone $\mathcal S^+_p:=\{\Sigma \in \mathcal S_p: x^\top \Sigma \thinspace x>0, x \in \mathbb R^p\backslash \{0\}\}$ \footnote{We use $A^T$ with a roman letter ``$^T$" as superscript to highlight dependence of the matrix $A$ on the tree $T$, and distinguish this from $A^{\sf T}$ to denote its transpose.} of symmetric positive definite matrices in $\mathbb R^{p \times p}$. For simplicity, we refer to strictly ultrametric matrices as ultrametric. 



Matrices in $\mathcal U_p$ are inverses of $M$-matrices \citep{AO}, and are used to model conditional dependencies in Gaussian models \citep{MTP2_Def_KARLIN1980, MTP2_InvM_KARLIN1983}. Associated with each ultrametric matrix in $\mathcal U_p$ is a rooted tree on $p$ labeled leaves.\footnote{Confusingly, in phylogenetics, rooted trees with leaves equidistant from the root are referred to as ultrametric trees with a corresponding distance matrix (between leaves), known as an ultrametric distance matrix. We do not consider such trees or the distance matrices.}. Such a tree $T$ encoded within an ultrametric matrix $\Sigma^T$ is not necessarily unique, but is helpful to interpret the following decomposition of an ultrametric matrix \citep[Theorem 2.2][]{Nabben1994}:
\begin{align}\label{eq:SigmaT_NaiveDecomp}
    \Sigma^T = \sum_{j=1}^{2p-1} d_j v_j v_j^\transp=VDV^\transp,
\end{align}
where $\{v_j\}$ are $p$-dimensional binary vectors with values in $\{0,1\}$ with $v_1=\bm 1$, the vector of ones, and $\{d_j\}$ are non-negative real numbers; there are $p$ vectors $v_j$ containing a single non-zero element with corresponding $d_j >0$. The matrix $V=(v_1,\ldots,v_{2p-1})^\top$ is known as the basis matrix of $\Sigma^T$ with the partition property \citep{MIP_Tr} that associates with a rooted tree structure \citep{Nabben1994}. Starting from the root of the tree pertaining to $v_1$, it encodes a recursive partition of $[p]$ such that every column $v_i$ with more than one non-zero elements there exists two other columns $v_j, v_k$ such that $v_i=v_j+v_k$. The matrix $V$ determines the topology of the tree $T$, while the diagonal matrix $D$ with entries ${d_j}$ stores the edge lengths. 

The set of $M$-matrices is a convex subset of $\mathcal S^+_p$ and this was used by \cite{SH} to compute the maximum likelihood estimate (MLE) of the (inverse) covariance matrix in a Gaussian graphical model. However, existing approaches to estimation on $\mathcal U_p$  are dominated by optimization methods based on projections or relaxations, since the inequalities in Definition 1 are non-convex. For example, \citet{MLE_MTP2} and \citet{10.1093/jjfinec/nbaa018} handle the inequalities by considering a dual problem. On the other hand, ultrametric matrices are a special case of linear covariance matrices introduced by \citet{TWA} and occur as covariance matrices of Brownian tree models introduced in phylogenetics \citep{Felsenstein}, and more generally as covariance matrices of marginal likelihoods in generative models for hierarchically correlated multivariate data \citep{DDT_Neal, yao2022probabilistic}. As a special case of linear covariance matrices, the MLE of $\Sigma^T$ can be computed given the basis matrix $V$ \citep{ZUR}; this corresponds to fixing the topology of the tree $T$ apriori, and the corresponding set of ultrametric matrices can be identified with a simplicial cone within a spectrahedron in $\mathcal S_p$ \citep{Sturmfels2021}. However, the basis matrix $V$ is part of the parameter space and needs to be estimated. 

Despite being a subset of the convex set of inverse $M$-matrices, the set $\mathcal U_p$ is neither convex nor a smooth manifold \citep{McCullagh2006StructuredCM}. However, to carry out sampling-based inference on $\mathcal U_p$ under a Bayesian setting, it is important to enable algorithms to make intrinsic local moves that do not leave the parameter space. Extrinsic geometries for $\mathcal U_p$ based on embeddings into $\mathcal S_p$ or $\mathcal S^{+}_{p}$ require projections to ensure that samples always assume values in $\mathcal U_p$ \citep[e.g.][]{MIP_Tr}, which may be inefficient owing to the non-trivial structure of $\cU_p$. 

\subsection{Rooted trees and representations}
Trees are acyclic graphs, and broadly, come in three flavors: rooted or unrooted; labeled, or unlabeled; binary or multifurcating. We focus on rooted, labeled, binary and multifurcating trees with edge lengths on $p+1$ leaves, described as follows. By tree, we mean a connected acyclic graph $T$ with a unique node of degree one refereed to as the root. Nodes with degree one are referred to as leaves and all other nodes have degree greater than two and are known as internal nodes. Leaves are labeled from the set $L=\{0,1,\ldots,p\}$, where 0 is the label of the leaf node connected to the root. All leaf edges have strictly positive length, excepting the \emph{root edge}  connecting to the root, which is non-negative. The topology of $T$ is its shape or combinatorial structure upon ignoring edge lengths. 

For the methodological development, internal nodes need no labeling, but their labeling helps with exposition in certain places, and they thus have labels from the set $\tilde L=\{u_1,\ldots,u_{p-2}\}$ following a depth-first scheme starting from the root \citep[e.g.][]{cormen2022introduction}; this induces a parent-child order on nodes that form edges in the set $\cE_T \subseteq (\tilde L \cup L) \times (\tilde L \cup L)$, which is the union of the set $\cE_T^I$ of edges connecting internal nodes and the set $\cE_T^L$ of edges connecting internal nodes to the $p+1$ leaves. The representation $e=(a,b) \in (\tilde L \cup L) \times (\tilde L \cup L) $ is referred to as the \emph{vertex representation} of the edge $e$. 

\emph{Resolved} trees $T$ are those with internal nodes of degree three and $(p-2)$ internal edges in $\cE^I_T$, while \emph{unresolved} trees are trees $T$ with fewer than $(p-2)$ internal edges containing internal nodes of degree four or higher. Thus, resolved trees contain either $(p-2)$ or $(p-1)$ internal nodes depending on whether the root edge if of length zero. 




\paragraph{Split representation and compatibility.} 
A representation of an internal edge in $\cE_T^I$ particularly useful to describe the topology of $T$ is based on the set of partitions into two of $L=\{0,1,\ldots,p\}$ called splits. More precisely, each edge $e \in \cE_T^I$ uniquely determines a split $L=A\cup A^c$ upon its removal from a tree $T$, where $A$ contains leaves on the descendant subtree of $e$ and its complement $A^c=L-A$ contains the rest of the leaves\footnote{Strictly speaking, as per terminology in phylogenetic literature, component containing the root is a clade, while the other is a split. However, we abuse terminology and refer to both as splits in order to be consistent with the terminology used in the work by \cite{Billera2001} that introduces the BHV space.}; thus the root edge determines the split $L=\{0\} \cup \{1,\ldots,p\}$. Denote by $e_A$ the corresponding edge with length $|e_A|$. For example, $e_{L\setminus \{0\}}$ is the root edge, and $|e_{L\setminus \{0\}}|$ its length. The set $A\subset L\setminus\{0\}$ identifies a split $L=A \cup A^c$, and we use split to refer to $A$ or the edge $e_A$ interchangeably; context will disambiguate the two. 

Arbitrary collections of splits do not characterize a valid tree topology, but only a collection of compatible ones do: two distinct edges $e_{A_1}$ and $e_{A_2}$ are compatible if exactly one of $A_1\cap A_2, A_1\cap A_2^c, A_1^c\cap A_2$ from the associated splits is empty set \citep[Theorem 3.1.4][]{semple2003phylogenetics}(upon noting $A_1^c\cap A_2^c$ always contains the root); identical edges are by default compatible. Again, we interchangeably refer to compatibility of splits $A_1$ and $A_2$ to sometimes mean compatibility of the edges $e_{A_1}$ and $e_{A_2}$, and this extends to a collection $\{A_1,\ldots,A_k\}$ of subsets of $L$. For example, we say $\{e_{A_1}, e_{A_2}\}$ is compatible with $\{e_{A_3}, e_{A_4}\}$ if any edge from the first set and any edge from the second set are compatible. Leaf edges $e_A \in \cE_T^L$, including the one from the root, associated with singleton splits $A \subset L$ are compatible with all internal edges in $\cE_T^I$, and thus do not contribute to the topology of $T$. A compatible edge set $\cE_T$ thus fully characterizes the topology of a tree $T$. There are $(2p-3)!!$ distinct topologies on fully resolved trees on $p+1$ leaves where the root is a leaf node labeled $0$. 

\begin{figure}
    \centering
    \includegraphics[width=\textwidth]{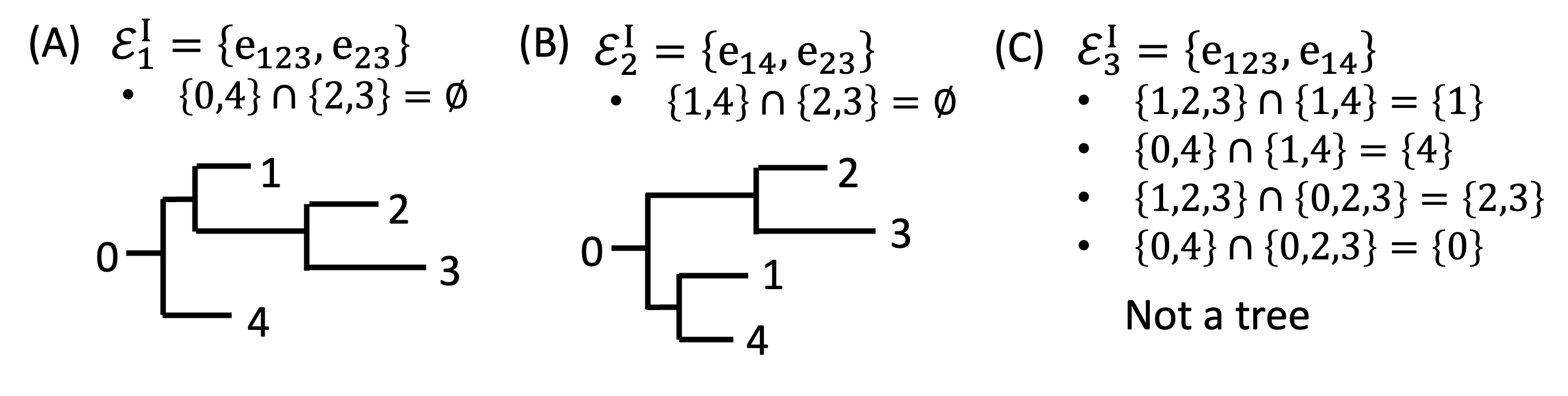}
    \caption{Three examples for demonstrating the compatibility of internal edge sets for $p=4$ with the corresponding tree topology. Two compatible sets are shown in Panel (A) and (B) with the concordant tree topology. Panel (C) exhibits incompatible edge set that characterizes no tree topology.}
    \label{fig:SpltComp}
\end{figure}

Three internal edge sets with $p=4$ are shown in Figure \ref{fig:SpltComp}. Panel (A) and (B) demonstrate two compatible internal edge sets. Two trees with different topology are characterized based on the corresponding compatible internal edge set. Contrarily, Panel (C) shows an incompatible edge set for none of the intersect is empty ($\{1,2,3\}\cap \{1,4\}, \{0,4\}\cap\{1,4\},\{1,2,3\}\cap\{0,2,3\}$ and $\{0,4\}\cap\{0,2,3\}$), which implies no tree topology can be characterized by the edge set.
 
Given an edge $e=(u_i,u_j)$ in the vertex representation, its edge representation $e_A$ is obtained by determining $A \subset L$ as the leaves of the unique subtree with edge $(u_i,u_j)$ connected to the root $u_i$ such that the descendant map on $e$ gives $\pi(e)=u_j$.

\subsection{Geometry of the BHV space for trees}
\label{sec:BHV}
The BHV space $\mathcal T^{I}_{p+1}$ parameterizes the space of rooted resolved and unresolved trees $T$ on $p+1$ leaves, where only the leaf nodes are labeled. It prescribes a continuous geometry based on the lengths $|e_A|$ of internal edges $e_A \in \cE_T^I$, where $A$ is associated with a split of $L$. A fully resolved topology is parameterized by $\mathbb R_{>0}^{p-2}$, where each axis corresponds to one of the $p-2$ internal splits that characterize the topology and the coordinates encode the corresponding lengths of the internal edges. The boundary of $\mathbb R_{>0}^{p-2}$ consists of unresolved trees with internal nodes of degree greater than 3, obtained by shrinking the internal edges to zero. Each of the $(2p-3)!!$ topologies is identified with a copy of $\mathbb R_{\geq 0}^{p-2}$, known as an orthant, and the BHV space $\mathcal T^{I}_{p+1}$ is defined by the $(2p-3)!!$ orthants glued isometrically along their common boundaries comprising unresolved trees. Panel (A) of Figure \ref{fig:TrCovDecomp_T}  illustrates that two neighboring orthants share a common edge in $\mathcal T_5^I$. Upon considering the lengths of $p+1$ leaf edges, we obtain the product space
\[
\mathcal T_{p+1}=\mathcal T^{I}_{p+1} \times \mathcal X_{p+1},
\]
where the second factor $\mathcal X_{p+1}:=\mathbb R_{ \geq0} \times \mathbb R^{p}_{>0}$ represents possible lengths of the edge connecting the root 0 and the strictly positive leaf edges unconnected to the root with labels in $[p]$. Denote by $\cL_T \in \mathcal X_{p+1}$. A tree $T$ is thus characterized by the pair $(\cE_{T}, \cL_T)$ of a compatible edge set and vector of edge lengths. The origin of $\mathcal T_{p+1}$ consists of trees with internal edges of length zero, a non-negative leaf edge connecting the root, and $p$ leaf edges of positive length. 

The distance $d_{\text{BHV}}(T_1,T_2)$ between two trees $T_1$ and $T_2$ on $p+1$ leaves is defined to be the infimum of lengths of paths between $T_1$ and $T_2$ in $\mathcal T^I_{p+1}$, which are straight lines within each orthant. It is known that the space $(\mathcal T^I_p, d_{\text{BHV}})$ is a geodesic metric space \citep{Billera2001}. A natural distance metric on $\mathcal T_{p+1}$ then is
\[
d_{\text{tree}}(T_1,T_2):=d_{\text{BHV}}(T_1,T_2)+\|x-y\|_2,
\]
where $x=(x_0,x_1,\ldots,x_p)^\transp, y=(y_0,y_1,\ldots,y_p)^\transp \in \mathcal X_{p+1}$ are the vectors of leaf edge lengths in $T_1$ and $T_2$, respectively.


\section{Stratified geometry of ultrametric matrices}\label{sec:geometry}
It is known that every ultrametric matrix  $\Sigma^T$ can be identified with \emph{some} tree $T$ \citep[][Theorem 3.16]{InvMMat_UltraMat}. Relatedly, based on the decomposition \eqref{eq:SigmaT_Decomp}, Theorem 2.2 of \cite{Nabben1994} also confirms the existence of a tree $T$, resolved or unresolved, within each ultrametric matrix $\Sigma^T$ via the decomposition \eqref{eq:SigmaT_Decomp}; this tree is not unique since, for example, permuting the columns of $V$ via $V \mapsto VP$ with a permutation matrix $P$ leaves $VDV^\top$ invariant. 

On the other hand, bijections between the set of rooted, leaf-labeled resolved trees with \emph{positive} internal edge lengths and the set $\tilde \cU_p$ of ultrametric matrices that are not necessarily strictly ultrametric have been established; see in \cite{semple2003phylogenetics} the bijection in Theorem 7.2.8 via the Farris transform of an ultrametric on a finite set. 

We demonstrate that by using the extended BHV space coordinates for each resolved/unresolved tree $T$ a bijection $\mathcal U_{p} \to \mathcal T_{p+1}$ may be constructed. This is achieved by demonstrating that redundancy due to permutations of the columns in the decomposition \eqref{eq:SigmaT_Decomp} is resolved when an ultrametric matrix is prescribed coordinates of the $\cT_{p+1}$; in other words, rooted trees on $p$ leaves encoded within the same ultrametric matrix $\Sigma^T$ map to the same point in $\cT_{p+1}$. 


\begin{theorem} [Bijection and pullback geometry on $\cU_p$]
\label{thm1}
There is a bijection $\Phi:\mathcal U_p \to \mathcal T_{p+1}$ such that when equipped with the metric 
\begin{align}\label{eq:matDist}
d(\Sigma^T_1,\Sigma^T_2):=d_{\textup{tree}}(\Phi(\Sigma_1^{T}),\Phi(\Sigma_2^{T})), 
\end{align}

the space $(\mathcal U_p, d)$ is a CAT(0) stratified geodesic metric space. 
\end{theorem}


Let $\Psi:=\Phi^{-1}$ be the inverse of $\Phi$. Theorem \ref{thm1} enables us to pullback the CAT(0) geometry from the tree space $\mathcal T_{p+1}$ onto the set $\mathcal U_p$ of ultrametric matrices using $\Psi$.  CAT(0) spaces are spaces with non-positive curvature in the Alexandrov sense, and corresponding geometry equips $\mathcal U_p$ with following properties that are particularly important in the sequel:
\begin{enumerate}
\item [(i)] There is a unique geodesic between any two ultrametric matrices that lies within $\mathcal U_p$, which enables us to develop a sampling algorithm that makes intrinsic local moves along geodesics in $\mathcal U_p$ based on `nearest neighbor interchanges' or `rotations' on $\cT_{p+1}$ \citep{Billera2001}; 
\item [(ii)]  Fr\'{e}chet means of probability measures exist and are unique; this enables computation of an estimate of the posterior mean using samples from the posterior distribution on $\mathcal U_p$. 
\end{enumerate}
For details on CAT(0) spaces, we refer to \citet{BH}. Additionally, the CAT(0) structure of  $\mathcal U_p$ opens up the possibility of using the plethora of statistical tools for the BHV space $\mathcal T^I_{p+1}$ (e.g., means and variances \citep{BO}, principal component analysis \citep{Nye2011}) and transfer results on to $\mathcal U_p$ with $\Psi$; most relevant to our work is computation of the geodesic distance $d$ on $\mathcal U_p$ using the polynomial-time algorithm for computing $d_{\text{BHV}}$ on $\mathcal T^I_{p+1}$ \citep{BHV_Dist} and fast computation of the Fr\'{e}chet mean \citep{FrechetMeanTr}. 


\subsection{How the strata are linked}
By pulling back the geometry of $\mathcal T_{p+1}$ via $\Psi=\Phi^{-1}$ a matrix representation of every tree $T$ in $\mathcal T_{p+1}$ is available, and is recorded in the following result, which appears to be new, and may be of independent interest; its proof follows from the proof of Theorem \ref{thm1}. 
\begin{corollary}[Split representation of an ultrametric matrix]
	\label{cor1}
	For every edge $e_A$ in a collection $\cE_T$ of compatible edges/splits corresponding to a tree $T$, there is a unique $p \times p$ symmetric binary matrix $E_{A}$ containing ones at all pairs of indices obtained from $A$ (the set of leaves on the descendent subtree of the edge $e_A$) and zeros elsewhere, such that the ultrametric matrix $\Sigma^T$ can be expressed in coordinates $(\cE_T, \cL_T)$ via the decomposition
	\begin{align}\label{eq:SigmaT_Decomp}
		\Sigma^T =\Psi((\cE_T, \cL_T)):=\sum_{e_A \in \cE_{T}} \lvert e_A \rvert E_{A}  \thickspace.
	\end{align}
\end{corollary}


For every $\Sigma^T$, the image $\Phi(\Sigma^T)$ is a tree with $p+1$ leaves. The preimage $\Psi(T)$ of every tree $T \in \mathcal T_{p+1}$ is a positive definite strictly ultrametric matrix in $\mathcal U_p$ (Lemma \ref{lem1} in Appendix). The preimage of each of the $(2p-3)!!$ orthants of dimension $(2p-1)$ in $\mathcal T_{p+1}$  is a stratum of $\mathcal U_p$. 
The ultrametric matrix $\Psi((\cE_T, \cL_T))=\sum_{e_{A}\in \cE_T} \lvert e_A \rvert E_{A}$ is located within a stratum identified by the vector $(E_{A_1},\ldots,E_{A_{K}})$ of binary rank-one matrices, where $K \leq 2p-1$, coordinatized by the point $(|e_1|,\ldots,|e_{2p-1}|)^\transp$ in $\mathbb R^{2p-1}_{\geq0}$. For convenience, we use the notation $\{E_A: e_A \in \cE_T\}$  to make explict the dependence of the binary matrices on the set of compatible splits, or just $\{E_A\}$ when the context is clear.

\begin{figure}[htb!]
    \includegraphics[width=\textwidth]{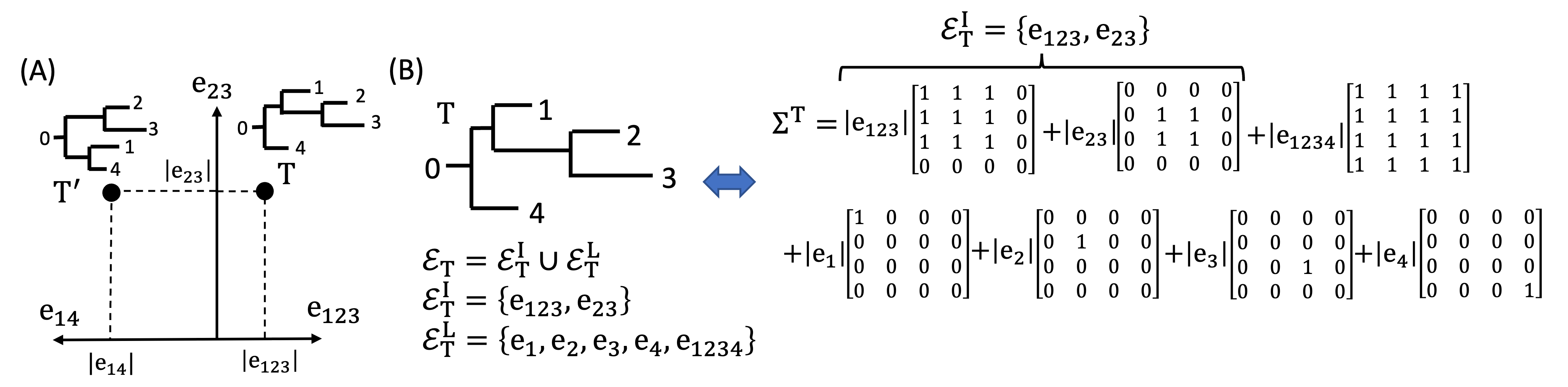}
    \caption{Bijection between the tree space $\mathcal T_5$ and the space $\mathcal U_4$ of ultrametric spaces in Theorem \ref{thm1} and the corresponding decomposition in Corollary \ref{cor1}: Panel (A) shows the BHV tree space $\mathcal T_5^I$ with axes identified with two compatible internal splits; Panel (B) shows how a rooted labeled tree in $\mathcal T_5$ uniquely identifies an ultrametric matrix $\Sigma^T$ in $\mathcal U_4$.}
    \label{fig:TrCovDecomp_T}
\end{figure}

 Figure \ref{fig:TrCovDecomp_T} illustrates the decomposition of an ultrametric matrix $\Sigma^T$  in $\mathcal U_4$ with corresponding tree $T \in \mathcal T_5$ and a compatible edge set $\cE_T=\{e_{123},e_{23},e_1,e_2,e_3,e_4\}$. 
For $I=\{1,\ldots,(2p-3)!!\}$, we introduce the notation 
\begin{equation}
\label{eq:stata}
\mathcal U_p=\bigcup\limits_{i \in I} \mathcal U^i_p
\end{equation}
to represent $\mathcal U_p$ as a disjoint union of $(2p-3)!!$ spaces $\mathcal U^i_p,i \in I$, each copy of $\mathbb R^{p-2}_{\geq 0} \times \mathcal X_{p+1}$ with common boundaries glued as described above. The origin common to each stratum $\mathcal U^i_p$ consists of diagonal matrices with positive entries along the diagonal. 

\begin{remark}
The representation of $\Sigma^T$ \eqref{eq:SigmaT_Decomp} may be compared to Corollary 2.3 of \cite{Sturmfels2021}, where their simplicial cone within the set of $p \times p$ symmetric matrices is spanned by the set $\{E_{A}\}$ that constitute the extremal rays and is identified with $\mathcal U^i_p$ in \eqref{eq:stata} for a particular $i$. 
\end{remark}

There are clearly many paths from the interior of a stratum to a codimension $s \geq 1$ boundary of $\mathcal U_p$; in other words, there are many copies of $\mathcal U_{q}$ within $\mathcal U_{p}$ for $q<p$. It is then natural to query:
How do ultrametric matrices on the boundaries `look'?  Are they sparser in some appropriate sense to those in the interior? To answer this in a relative sense, note that boundaries in the BHV space are identified (glued) isometrically along their common faces. Pulling this structure back onto $\mathcal U_p$ engenders a similar identification of the boundaries of $\mathcal U_p$. 

Recall that there are $(2k+1)!!$ binary trees with $k+2$ leaves. With $\textsf{vec}: \mathbb R^{p \times p} \to \mathbb R^{p^2}$ as the vectorization map, consider the half-vectorization map $\textsf{vech}(x):=A \textsf{vec}(x)$,
where the matrix 
\[
A:=\sum_{i \geq j}(u_{ij}\otimes e_j^\transp \otimes e^\transp _i) \in \mathbb R^{p(p+1)/2 \times p^2}
\]
picks out the lower triangular part of the vectorization, $\{e_j\}$ is the standard basis in $\mathbb R^{p^2}$,
and $u_{ij}$ is a $p(p+1)/2$-dimensional unit vector with 1 in position $(j-1)p+i-j(j-1)/2$ and 0 elsewhere. 
Consider a partial order $\preceq$ on $\mathcal U_p$ defined as $A \preceq B$ if $\textsf{vech}(A) \leq \textsf{vech}(B)$, and $\bm x \leq \bm y$ on $\mathbb R^{p(p+1)/2}$ if every element of $\bm x$ is utmost its corresponding element in $\bm y$. 
\begin{proposition}
\label{prop1}
For $1<k \leq (p-2) $, an ultrametric matrix $\Sigma^T$ on a $(2p-1-k)$-dimensional boundary can be reached along geodesics from distinct ultrametric matrices lying in the interior of $(2k+1)!!$ strata such that $\Sigma^T$ is smaller with respect to $\preceq$ to each of the $(2k+1)!!$ ultrametric matrices. Moreover, an ultrametric matrix on a stratum of dimension $j$ is smaller with respect to $\preceq$ than any other on a stratum of dimension $k$, where $j<k$. 
\end{proposition}

\begin{remark}
\label{rmk_embedReal}
Coefficients in the sum \eqref{eq:SigmaT_Decomp} are lengths of edges in the set $\cE_T$ of edges given by compatible splits of $L$. There are a total of $N=2^{p}-2$ possible nonempty splits of $L$ of which $N-p$ correspond to the internal edges. Thus, using $\Phi$, the set $\mathcal U_p$ can be given extrinsic coordinates via an embedding of $\mathcal T_{p+1}$ into $\mathbb R^{N-p}$ where each set $\{|e_A|, e_A \in \cE_T \}$ of edge lengths can be identified with a vector in $\mathbb R^{N-p}$, obtained upon choosing an ordering of the $N-p$ splits; the $j$th split is associated with the $j$th standard basis vector consisting of a one in the $j$th position and zero everywhere else. Such an embedding results in an alternative representation of $\Sigma^T$ as very sparse vector in $\mathbb R^{N-p}$, since only a small subset of vectors correspond to compatible splits. The geometric picture of $\mathcal U_p$ hence obtained with such an embedding is not transparent. 
\end{remark}

\section{Computing the bijection $\Phi$}
\label{sec:compute_bijection}
\begin{figure}[!htb]
    \centering
    \includegraphics[width=\textwidth]{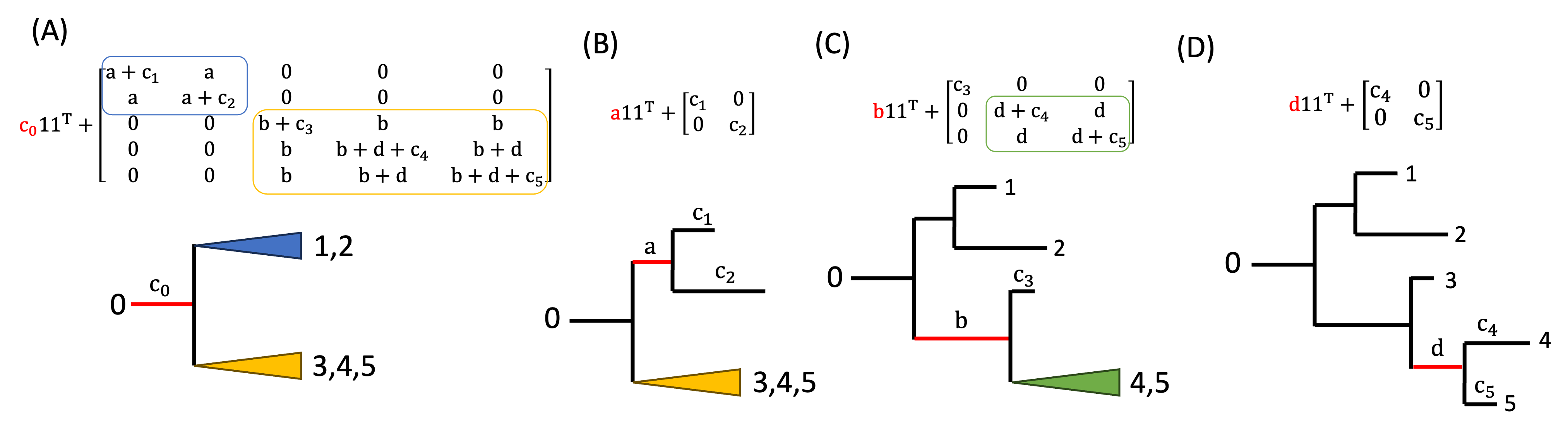}
    \caption{Panels (A)-(D) demonstrate how the recursive decomposition of $\Sigma^T$ for a resolved tree $T$ with $p=5$ leaves. Colored triangles in the trees represent subtrees corresponding to subsets of leaves.}
    \label{fig:Mat2Tree}
\end{figure}

We now describe how $\Phi(\Sigma^T)$ may be computed for a given ultrametric matrix $\Sigma^T$ by identifying the coordinates in $\mathcal T_{p+1}$ given by the decomposition \eqref{eq:SigmaT_Decomp}. Let $\alpha \geq 0$ be the smallest element of $\Sigma^T$. 
The decomposition \eqref{eq:SigmaT_NaiveDecomp} is based on the fact that there exists a positive integer $r$ with $r=1,\ldots,p-1$ and a $p \times p$ permutation matrix $P$ such that the matrix $\tilde \Sigma^T=\Sigma^T-\alpha \bm 1 \bm 1^\transp$ can be expressed as
\begin{equation}
\label{eq:perm}
P\tilde \Sigma^T P^\transp=
\begin{pmatrix}
C_1&0\\
0&C_2\\
\end{pmatrix},
\end{equation}
where $C_1$ and $C_2$ are $r \times r$ and $(p-r) \times (p-r)$ ultrametric matrices, respectively \citep[][Proposition 2.1]{Nabben1994}. Here $\alpha=d_1$ in \eqref{eq:SigmaT_NaiveDecomp}, and is the length of root edge, and $C_1$ and $C_2$ are the left and right subtrees. The result can be applied recursively to the matrices $C_1-\alpha_1\bm 1 \bm 1^\transp$ and $C_2-\alpha_2\bm 1 \bm 1^\transp$, where $\alpha_1$ and $\alpha_2$ are respectively the smallest elements of $C_1$ and $C_2$, until each block is a positive scalar. Figure \ref{fig:Mat2Tree} provides an illustration for a resolved tree $T$ with $p=5$ leaves. 

For unresolved trees, the number of subtrees at each node, starting from root, is at least two, and this structure is latent within the ultrametric matrices $C_1$ and $C_2$ at each stage of the recursion. From the proof of Proposition 2.1 of \cite{Nabben1994} it is evident that the result may be refined to 
\begin{equation}
\label{eq:perm2}
P\tilde \Sigma^T P^\transp=
\begin{pmatrix}
C_1&0&\cdots&0\\
0&C_2&\cdots &0\\
\vdots & \vdots &\ddots & \vdots\\
0&0 &\cdots &C_{k}
\end{pmatrix},
\end{equation}
to account for the case when the root has $k$ subtrees, where $C_i$ for each $i=1,\ldots,k$ is an $r_i \times r_i$ ultrametric matrix, with $r_1+\cdots+r_{k}=p$, and $P$ is a $p \times p$ permutation matrix. Each $C_i$ corresponds to a subtree $T_i$ with the same root node as other subtrees $T_j$ corresponding to $C_j$ with $j \neq i$. The dimension of $C_i$ determines the number of leaves of the subtree $T_i$, and the smallest element of $C_i$ is the length of the edge connecting to the root node of $T_i$. This process may be continued recursively for each ultrametric $C_i$ until all diagonal blocks are of dimension $1 \times 1$ to obtain the decomposition \eqref{eq:SigmaT_Decomp}.
\begin{algorithm}[!htb]
\hrulefill
	\small 
	\caption{Computing the representation in \eqref{eq:perm2} for an ultrametric matrix}
    \label{algo:bijection}
	\SetKwInOut{Input}{Input}
	\SetKwInOut{Output}{Output}
    \SetKwRepeat{Do}{do}{While}
	\SetKwFunction{proc}{}
	\texttt{
	\Input{$p \times p$ ultrametric matrix $\Sigma^T=\{\Sigma^T_{ij}\}$ with $p\geq2$}
	\Output{\vspace*{-5pt}
		\begin{itemize}
			\itemsep -0.5em
			\item[-] $p \times p$ permutation matrix $P$ and block-wise diagonal matrix $\textrm{diag}(C_1,\ldots,C_{k})$ in \eqref{eq:perm2};
            \item[-] Edge lengths $\{|e_A|\}$ and matrices $\{E_A\}$ of subtree at child node of the root.
		\end{itemize}%
    }
    \While{$p\geq 2$}{
        \nl Compute $\tilde \Sigma^T=\Sigma^T-\alpha \bm 1 \bm 1^\transp$ with $\alpha=\min\{\Sigma^T_{ij}\}$ as the length of leaf edge $(0,u_1)$\;
        \For{$i=1,\ldots,p$} 
        {
            \uIf{$\tilde{\Sigma}^T_{ij}>0, i\neq j$}{
			   Let $i,j \in A_r$ so that leaves $i$ and $j$ are part of the same subtree\;
            }
			\Else{
				Leaves $i$ and $j$ are in different subtrees with $i\in A_r, j\in A_s$ with $A_r \cap A_s=\emptyset $\;
            }
        }
        \nl For $r=1,\ldots,k$, compute $a_r= \min A_r$ and relabel $A_r$ in increasing order of $a_1,\ldots,a_k$ to obtain $A_{r_1},\ldots,A_{r_k}$ with $[p]=A_{r_1} \cup \cdots \cup A_{r_k}$\;
        \nl For each $j=1,\ldots,k$, reorder elements of $A_{r_j}$ in increasing order, and concatenate them to generate a permutation $\sigma:[p] \to [p]$\;
        \For{$i=1,\ldots,p$}{
            Assign $P_{i,\sigma(i)}=1$ and zeros to the rest\;
        }
        \nl Return block-wise matrix $\textrm{diag}(C_1,\ldots,C_k)=P\tilde{\Sigma}^T P^\transp$\;
        \For{$l=1,\ldots,k$, construct $E_{A_{r_l}}$}{
            \uIf{$\tilde{\Sigma}^T_{ij}>0, i\neq j$}{
                set $(i,j)$ element of $E_{A_{r_l}}$ to 1 if $i\in A_{r_l}$\;
            }
			\Else{
				set $(i,j)$ element of $E_{A_{r_l}}$ to 0\;
            }
            \nl Assign smallest element of $C_j$ as edge length $\lvert e_{A_{r_j}}\rvert$\;
        }
        \nl Return $\{E_{A_{r_1}},\ldots,E_{A_{r_k}}\}$ and corresponding $\{\lvert e_{A_{r_1}}\rvert,\ldots,\lvert e_{A_{r_k}}\rvert\}$\;
    }
    }
\hrulefill
\end{algorithm}

The key observation is that the ultrametric matrices $C_1,\ldots,C_{k}$ may be determined by grouping together the columns of $\tilde \Sigma^T$ which have: (a) the same pattern of non-zero elements; (b) the same smallest element. The number of groups formed determines the number $k$ of children of the node connected to the root of $T$. This property is noted in equation (1) of \citet{HoAne2014}, but information on the permutation matrix $P$ is not available in their work. The permutation $P$ in \eqref{eq:perm2} is not unique since the labeling/ordering of the groups, and ordering of the columns of $\tilde \Sigma^T$ within each of the $k$ groups is arbitrary. Proof of Theorem \ref{thm1} clarifies how this indeterminacy is inconsequential for the bijective property of $\Phi$; the same argument applies to the permutation matrix, say, $P_j$ of appropriate dimension at recursion stage $j$. In practice, however, the bijection $\Phi$ is constructed by determining the permutation each stage. 

\begin{figure}
        \centering
        \includegraphics[width=\linewidth]{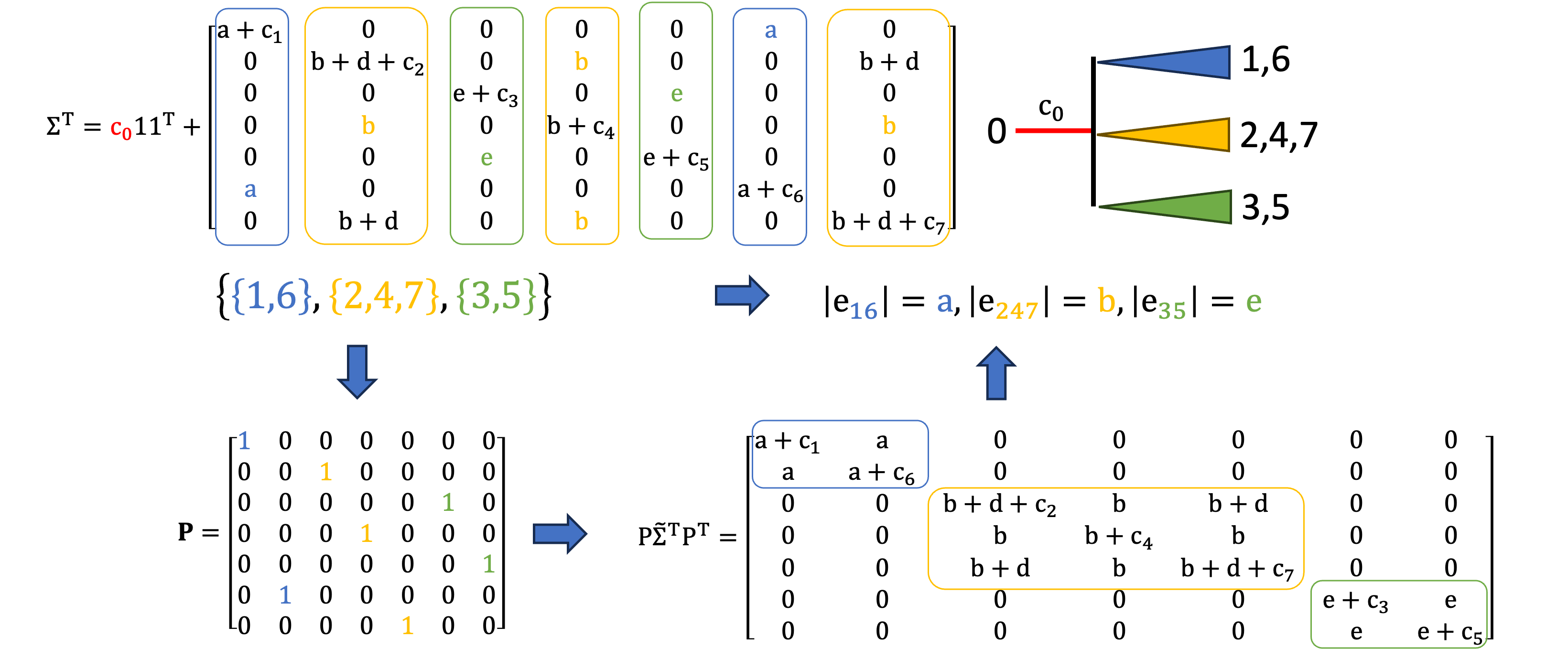}
        \caption{\small Algorithm \ref{algo:bijection} to compute the representation in \eqref{eq:perm2} for an unresolved tree on $p=7$ leaves. Here $k=3$, and the tree contains a single internal edge of length zero with $\Sigma^T$ thus lying on a co-dimension one common boundary of three strata within $\mathcal U_p$. Repeated use of the Algorithm \ref{algo:bijection} on each of the three block diagonal matrices provides the bijection $\Phi$.}
        \label{fig:alg_Mat2Tr}
    \end{figure}

Algorithm \ref{algo:bijection} prescribes a procedure to compute a permutation $P$ in \eqref{eq:perm2} that identifies the $k$ subtrees $T_1,\ldots,T_k$ corresponding to the ultrametric matrices $C_1,\ldots,C_k$, and use it to construct the edge lengths $\{|e_{A_{r_j}}|\}$ and the basis matrices $\{E_{A_{r_j}}\}$ of each subtree $T_j$, where $r_1,\ldots,r_k$ is a particular ordering of $1,\ldots,k$. Step 2 in Algorithm \ref{algo:bijection} represents a particular choice of ordering of the groups, while step 3 represents a specific ordering within each group, which together determines the permutation $\sigma:[p] \to [p]$ used to construct $P$. The bijection $\Phi$ is computed by reapplying Algorithm \ref{algo:bijection} to all other subtrees of $T_1,\ldots,T_k$ to obtain their edge lengths and basis matrices (recursive computation of \eqref{eq:perm2}), until the full set of edge lengths $\{|e_{A}|\}$ and basis matrices $\{|E_{A}|\}$ is obtained. Both $\{|e_{A}|\}$ and $\{|E_{A}|\}$ contain $K$ elements corresponding to the tree $T$, where $K=(q+p) \leq 2p-1$ and $q$ is the number of non-zero internal edges or the root edge.
Figure \ref{fig:alg_Mat2Tr} offers an example for an ultrametric matrix $\Sigma^T$ with $k=3$ in its representation in \eqref{eq:perm2}, with an unresolved tree $T$ on a codimension one boundary shared by three orthants in $\mathcal T_{p+1}$ on $p=7$ leaves.

\section{Prior distributions on ultrametric matrices}\label{sec:TrPrior}
From Theorem \ref{thm1} and Corollary \ref{cor1}, we note that an ultrametric matrix $\Sigma^T  \in \mathcal U_p$ can be given coordinates $(\cE_T,\cL_T)$ corresponding to the tree $T$ via $\Psi$ . We can thus define a prior distribution on $\mathcal U_p$ pushing forward one on $\mathcal T_{p+1}$ under $\Psi$. Consider the product measure $\text{d}\eta:=\text{d}\mathbb N\text{d}\bm x$, where $\text{d}\mathbb N$ is the counting measure on $\mathbb N:=\{0\} \cup \{1,2,\ldots\}$ and $\text{d}\bm x$ is the Lebesgue measure on $\mathbb R_{>0}^{p-2} \times \mathcal X_{p+1}$. Consider the probability distribution $\mu=\pi\text{d}\eta$ on $\mathcal T_{p+1}$ with density $\pi:=\pi_\cE \thickspace \pi_{\cL|\cE}$ that factors into a density $\pi_\cE$ on sets of compatible splits and a density $ \pi_{\cL|\cE}$ on edge lengths conditional on the splits; $\pi$ can always be chosen to be a valid probability density since $\int_{\mathcal T_{p+1}} \pi \text{d}\eta<\infty$ by the CAT(0) property of $\mathcal T_{p+1}$. Define the distribution 
\[
\nu(A)=\mu \circ \Phi(A), \quad A \subset \mathcal U_{p},
\]
on $\mathcal U_p$. The distribution $\nu$ need not be absolutely continuous with respect to $\mu$ and may hence not have a density. Under this setup, we specify a prior distribution $\nu$ by first doing so on $\mathcal T_{p+1}$ via the density $\pi$. 

There are a multitude of evolutionary probability models in phylogenetics which may used to define the prior $\pi$; see \cite{JF}, and \cite{paradis2006analysis} for a computational perspective. For the density $\pi_\cE$ on the splits, we consider the Gibbs-type fragmentation model \citep{GibbsFragTr} such that for a partition $\{A_1,\ldots,A_k\}$ of $B \subseteq \cE_T$ with  $k \geq 2$, 
\begin{align}\label{eq:genFrag}
	\pi_\cE(\cE_T)=\prod_{e_{A_1},\ldots,e_{A_k}} \pi_{\text{split}}(e_{A_1}, \ldots, e_{A_k} \mid B),
\end{align}
where $\pi_{\text{split}}(e_{A_1}, \ldots, e_{A_k} \mid B)$ characterizes the chance that the subtree determined by $B$ has blocks $A_1,\ldots,A_k$. The model is Markovian and consistent:  $k$ subtrees restricted to $A_{1},\ldots,A_{k}$ are independently distributed as $\pi_{\cE_{T_{A_{1}}}}, \ldots,\pi_{\cE_{T_{A_{k}}}}$. The density $\pi_\cE$ is assumed to be exchangeable so that it depends only on the numbers of elements in $A_1,\ldots,A_k$ \citep{betaSplit}. Other options include the restricted-exchangeable \citep{chen2013restricted} and non-exchangeable \citep{alpha-gamma-model} models of fragmentation. Although not general enough to handle multifurcating trees, for binary trees with $k=2$, models that additionally use a conditional clade probability \citep[e.g.][]{Larget2013}, or those which consider a more flexible structure to account for ancestors of $e_{A_1}$ and $e_{A_2}$ instead of only the immediate parent $e_{B}$ \citep[e.g.][]{Zhang2018}, may also be considered. However, keeping track of leaf labels when using models that are not exchangeable may significantly increase computational costs during inference. 

For resolved binary trees with $k=2$, \citet{Berestycki2007} considered a time-irreversible Markovian fragmentation process in order to define the splitting function $\pi_{\text{split}}$, while \citet{Pitman2006} defined a Gibbs-type alternative which defines $\pi_{\text{split}}$ as a product of weights that depend only on the size of sub-blocks. For such a Gibbs-type alternative, Theorem 2 of \citet{GibbsFragTr} showed that the beta-splitting model of \citet{betaSplit} is the only one that results in a consistent Markovian binary fragmentation-based prior $\pi_\cE$ on topology. The beta-splitting model prescribes the specification
\begin{align}\label{eq:betaSplt}
	\pi_{\text{split}}(e_A, e_B \mid e_{A\cup B}) \propto \frac{\Gamma(n_A + \beta+ 1)\Gamma(n_B + \beta+ 1)}{\Gamma(n_A + n_B + 2\beta+2)},
\end{align}
where $n_A$ is the cardinality of the set $A$ and $\beta \in (-2,\infty]$ is the hyper-parameter. For example, $\beta=-1.5$ leads to the uniform density $\frac{1}{(2p-3)!!}$ on topology, while $\beta=0$ corresponds to the Yule model \citep{YuleModel}. 

For unresolved trees containing multifurcating nodes, a Poisson-Dirichlet model based on multifarcating Gibbs fragmentations can be used to define a consistent Markovian prior $\pi_\cE$ that generalizes the beta-splitting model.  The Poisson-Dirichlet model contains an additional hyperparameter $\theta$ and $\alpha=2\beta/3$, which is a reparametrization of $\beta$, such that the multifurcating structure is determined by values of the tuple $(\theta,\alpha)$ \citep[Theorem 8]{GibbsFragTr}. For example, $\theta=-m\alpha$ and $\alpha<0$ results in a multifurcating tree with at most $m$ descendants, and when $m=2$ the beta-splitting model for binary trees is recovered. We consider the case of $\theta>-2\alpha$ and $0\leq\alpha<1$ with $(\theta,\alpha)=(1,0)$ that allows a flexible prior with an arbitrary number of descendants.

We assume that the density $ \pi_{\cL|\cE}$ depends on $\cE$ only through the number of non-zero edge lengths, and that the joint distribution of $q$ edge lengths becomes a product of $q$ one-dimensional marginals on $\mathbb R_{>0}$, where $q=p,\ldots,2p-1$. The Lebesgue component $\text{d}\bm x$ of the dominating measure $\text{d}\mu$ can be viewed as a mixture $\sum_{j=p}^{q}\text{d}\bm x_j $ decomposed over the different strata upon identifying the boundaries of the individual strata of the same dimension.

\section{Posterior inference under a Gaussian latent tree model}\label{sec:Inference}
We consider two algorithms to sample from the posterior distribution on $\mathcal U_p$ given a Bayesian model with mean-zero Gaussian likelihood on random vectors. They are both adaptations of existing algorithms from Bayesian phylogenetics utlizing the connection to the extended treespace $\mathcal T_{p+1}$:
\begin{enumerate}
\item [(i)] A Hamiltonian Monte Carlo (HMC) algorithm on the interior of $\mathcal U_p$ for binary trees, which adapts one proposed by  \cite{ppHMC_Dinh2017} on $\mathcal T_{p+1}$; 
\item [(ii)] a Metropolis-Hastings (MH) algorithm for a reparameterized model on $\mathcal T_{p+1}$, for binary trees on the interior of the orthants and multifurcating trees of boundaries of adjacent orthants, based on nearest neighbour interchange (NNI) moves, which corresponds to moves along geodesics on the pullback geometry on $\mathcal U_p$.
\end{enumerate}
We focus on the MH algorithm in the main paper, and have provided details for the HMC in the Supplementary S1. 

\subsection{Sampling on $\mathcal T_{p+1}$ in phylogenetics}
Bayesian inferential methods for phylogenetic trees is a highly active and well-developed research area; see, for example, the \citet{mau1999bayesian} for an overview of the state-of-the-art 25 years ago highlighting issues that are still relevant today,  and \citet{barido2024practical} for a more recent overview; see the popular \texttt{MrBayes}\footnote{\url{https://nbisweden.github.io/MrBayes/}} for ready-to-use computational tools; \citet{zhang2024variational} for variational approaches; and \citet{hohna2024sequential} for sequential methods. 

Sampling methods for multifurcating trees (polytomies) have been considered in phylogenetics literature \citep[e.g.][]{Whelan2010, Lin2011}, and are classified as ``soft'' when multifurcation is due to poor resolution of the underlying binary tree, or ``hard'' when multiple multifurcations occur simultaneously \citep{Walsh1999}. 
We focus on the hard variety, for which there are two types of methods: (i) collapsing edges to create multifurcating nodes under a fixed topology \citep[e.g.][]{Zhang2021}; (ii) proposing a new multifurcating tree without assuming a fixed topology \citep[e.g.][]{Lewis2005,Lewis2015,PYDT}. Methods for the former may be inefficient if the topology considered is far away from the true one, while methods latter are typically computationally intensive requiring complex tree operations to propose a new tree with reversible-jump MCMC (except for \citet{PYDT}). 

\citet{Lakner2008} classify the different types of tree proposals within an MCMC setting as branch-change and branch-rearrangement; in the former, continuous edges/branches are updated to make moves which may or may not result in a change to tree topology, while in the latter moves are made by pruning and re-grafting subtrees with mandatory changes to tree topology. Within the geometry of the BHV space, moving along the unique geodesic in $\mathcal T_{p+1}$ between trees $T_1$ and $T_2$ may be viewed as combination of the branch-change and branch-reaarrangement methods: if $T_1$ and $T_2$ have the same topology, then the vector of edge lengths of $T_1$ is linearly updated to reach $T_2$; if $T_1$ and $T_2$ have different topologies but share some internal edges, then the geodesic is determined by nearest-neighbor interchange (NNI) moves obtained by shrinking some internal edges to zero, moving to the boundaries of the strata, and the moving out to new strata by growing internal edges. In particular, if $T_1$ and $T_2$ are trees that differ in a single internal edge, the geodesic is determined by a single NNI, and the distance between $T_1$ and $T_2$ is the length of the geodesic.

\subsection{MH-based posterior sampling}
The algorithm uses the bijection $\Phi$ to sample from the posterior of $\mathcal U_{p+1}$ by exploiting the existing sampling tools on $\mathcal T_{p+1}$. We consider the Gaussian latent tree model $\big\{N_p(\bm 0, \Sigma^T): \Sigma^T \in \mathcal U_p\big\}$ for multivariate data where correlations between components of a random vector is modeled through a tree $T$. The model is a special case of covariance graph models \citep{CW1}, which has received considerable attention \citep[e.g.,][]{CTAW, SZAS, PZ, LD}, however, mostly restricted to estimation strategies. 

Using the prior distribution from Section \ref{sec:TrPrior}, we consider the following Bayesian model:
\begin{align*}
    \bX_1, \ldots, \bX_n \thinspace | \thinspace \Sigma^T &\overset{\text{i.i.d.}}{\sim} N_p(\bm 0,\Sigma^T);\\
    \Sigma^T &\sim \nu ,
\end{align*}
where $\nu=\mu \circ \Phi=\pi \text{d}\eta \circ \Phi$. The density $\pi$ is chosen with $\pi_\cE$ corresponding to the density \eqref{eq:genFrag} with $\beta=-1.5$ for the binary case and $(\theta,\alpha)=(1,0)$ for the multifurcating case. 
$\pi_{\cL|\cE}$ is taken to to be a product of $q$ one-dimensional exponential densities $\textrm{Exp}(a)$ with a common mean $a$, where $q=p,\ldots,2p-1$ depending on the dimension of the stratum the tree $T$ with split set $\cE$ lies in. Under the beta-splitting model with $\beta=-1.5$, the dimension $q=2p-1$ or $2p-2$ depending on whether the leaf edge connecting the root is of non-zero length, and the prior puts zero mass on the boundaries of the strata. Extensions to non-Gaussian likelihoods (e.g., multivariate $t$-distribution and elliptical distributions) and presence of additional parameters (e.g., mean), can be carried out along similar lines once the advantages in the geometry-driven approach to sampling for the Gaussian case is well-understood.  

The bijection $\Phi$ in Theorem \ref{thm1} enables us to consider the equivalent reparameterized model 
\begin{align}
\label{eq:normModel}
 \bX_1, \ldots, \bX_n \thinspace | \thinspace \Psi(T) &\overset{\text{i.i.d.}}{\sim}N_p(\bm 0,\Psi(T));\\
 T & \sim \mu\thinspace \nonumber,
\end{align}
which is the model we use in what follows. We choose the prior $\mu$ on $\mathcal T_{p+1}$ to mainly demonstrate benefits of the uncovered bijection $\Phi$ for developing an efficient sampler for the posterior distribution $\nu(\cdot|\bX_1, \ldots, \bX_n)$ on $\mathcal U_p$ by exploiting existing sampling algorithms for trees in $\mathcal T_{p+1}$.

Although the pullback geometry of $\mathcal U_{p}$ is not explicitly used in the algorithms, the implicit geometry leads to two important inferential implications:
\begin{enumerate}
 \item [(a)] Sampling moves along geodesics ensures that projections are not required \citep[e.g.][]{MIP_Tr}, and leads to better exploration of $\mathcal U_p$;
    \item [(b)] use of the intrinsic metric $d$ on $\mathcal U_p$ enables computing statistical summaries of posterior samples, such as the  Fr\'echet mean, and modes of variability via PCA on $\mathcal T_{p+1}$ \citep{Nye2011}.
\end{enumerate}

Specification of the density $\pi$ on $\mathcal T_{p+1}$ was facilitated by the decomposition \eqref{eq:SigmaT_Decomp} of an ultrametric metric into its constituent topological component $\{E_A\}$ of binary matrices which encode information of the set of compatible splits $\cE_{T}$, and geometric component $\cL_T$ consisting of edge lengths. 

Denote by $\Sigma^{T_{(l)}}=\Psi(T_{(l)}) \in \mathcal U_p$ the $l$th iterate of an algorithm to sample from the posterior distribution $\nu(\Sigma^T|\bX_1,\ldots,\bX_n)$. More generally, an $(l)$ in either the subscript or superscript of any quantity is used to denote its value at the $l$th iteration. The algorithm moves along the geodesic starting at $\Sigma^{T_{(l)}}$ towards a point in an adjacent stratum, chosen with uniform probability, by making a NNI move on $\mathcal T_{p+1}$. 




First consider the algorithm for resolved trees with respect to the beta-splitting prior $\pi$ on $\mathcal T_{p+1}$ with no mass on unresolved trees. Suppose that iterate $\Sigma^{T_{(l)}}=\Psi(T_{(l)})$ lies in stratum $\mathcal U^i_p$ for some $i$:
\begin{enumerate}
	\item[(i)] 	Compute $\Phi(\Sigma^{T_{(l)}})=(\cE_{T_{(l)}},  \cL_{T_{(l)}})$, and determine the orthant of $\mathcal T_{p+1}$ the tree $T_{(l)}$ lies in;
	\item [(ii)] Randomly choose an internal edge/split $e_A \in \cE_{T_{(l)}}^I$ and set $|e_A|=0$ so that $T_{(l)}$ now lies on a codimension one boundary of $\mathcal T_{p+1}$; identify,  via the corresponding changes to $\cE_{T_{(l)}}$, the two nearest neighbors, distinct orthants $j$ and $k$ of $\mathcal T_{p+1}$ with $j\neq i$ and $k \neq i$, of $T_{(l)}$ there exist two and only two neighboring orthants distinct from the one where $T_{(l)}$ lies; see Proposition \ref{prop1} with $k=1$);
	\item[(iii)] Randomly choose between the two adjacent orthants $j$ or $k$  of $\mathcal T_{p+1}$ to decide on the updated topology $\cE_{T_{(l+1)}}$ of $T_{(l+1)}$;
	\item [(iv)] Update $\cL_{T_{(l)}}$ to  $\cL_{T_{(l+1)}}$ to obtain $T_{(l+1)}=(\cE_{T_{(l+1)}}, \cL_{T_{(l+1)}})$;
	\item [(v)] Compute $\Psi({T_{(l+1)}})=\Sigma^{T_{(l+1)}}$ using Algorithm \ref{algo:bijection}.
\end{enumerate}
If the topology update in steps (ii) and (iii) is rejected, then a local move constitutes moving along a geodesic within the same orthant by merely updating edge lengths. The motivation behind making a mandatory change in topology (which indeed can be rejected) is to make many computationally cheap local moves over a large number of iterations rather than a few computationally expensive moves, resulting in better exploration of the space of rooted trees. 

The algorithm above may be modified to enable sampling of unresolved trees on orthant boundaries by choosing a Poisson-Dirichlet prior $\pi$ for suitable values of $\theta$ and $\alpha$, and replacing steps (ii) and (iii) of the above algorithm as follows. 
\begin{enumerate}
	\item [(ii')] Randomly choose an internal edge/split $e_A \in \cE_{T_{(l)}}^I$ and set $|e_A|=0$ so that $T_{(l)}$ is lies in an orthant of codimension $k \geq 1$, which shares its boundary with $(2k+1)!!$ adjacent orthants (Proposition \ref{prop1});
	\item[(iii')] Excluding the orthant $T_{(l)}$ is in, choose with equal probability of $\frac{1}{(2k+1)!!}$ to stay on the boundary or move to any of the adjacent orthants to updated topology $\cE_{T_{(l+1)}}$ of $T_{(l+1)}$.
\end{enumerate}

\begin{figure}[htb!]
    \centering
    \includegraphics[width=\textwidth]{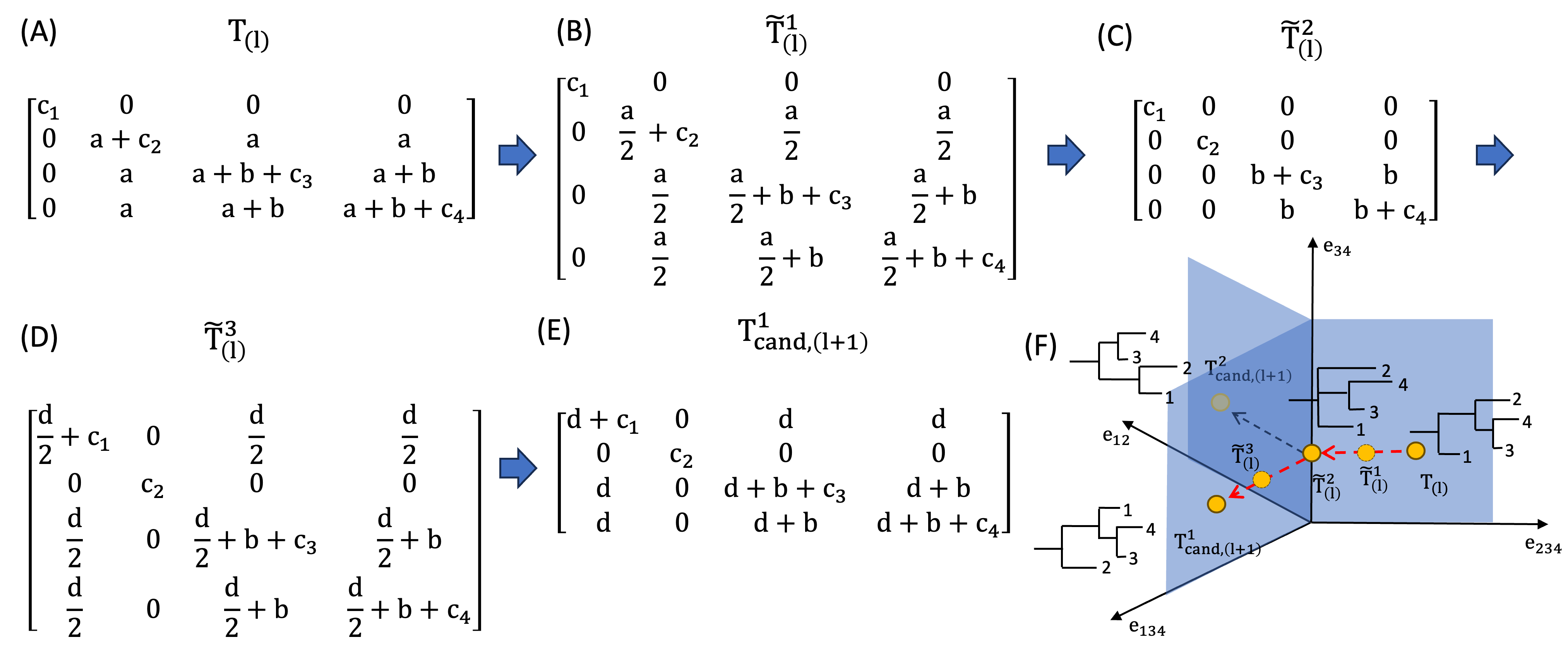}
    \caption{Updating topology of a resolved tree in $\mathcal T^I_5$ using a nearest neighbour interchange (NNI) step in Algorithm \ref{algo_MH}: Given tree $T_{(l)}$ at the $l$th iteration in orthant $i$, the proposal function randomly shrinks an internal edge and moves to an intermediate unresolved tree $\tilde{T}_{(l)}^2$ on boundary common to two other distinct orthants $j$ and $k$ with two candidate trees $T^1_{{\sf cand},(l+1)}$ and $T^2_{{\sf cand},(l+1)}$, one of which is chosen with equal probability, and reached by following geodesics (red and black dashed lines) from $T_{(l)}$. Panels (A)-(E) show three ultrametric matrices along the geodesic connecting $\Psi(T_{(l)})$ to candidate $\Psi(T_{{\sf cand},(l+1)}^1)$ (red dashed line). }

    \label{fig:BHV_localMove}
\end{figure}

To illustrate the local moves between resolved trees when using a beta-splitting prior $\pi$ with no mass on the orthant boundaries, we detail the update step in the space $\mathcal T_5$ of four-leaf trees in Figure \ref{fig:BHV_localMove}. Given a tree $T_{(l)}$ with four leaves at the $l$th iteration, we propose a candidate via a geodesic move from $T_{(l)}$ to $T^1_{{\sf cand},(l+1)}$ with every matrix on the path being ultrametric; this is highlighted by a path (red dashed line) in panel (F). Five matrices in $\mathcal U_4$ on the geodesic path from $T_{(l)}$ to $T_{{\sf cand},(l+1)}^1$ are shown in panels (A)-(E). 
The above algorithm is a modification of the one proposed by \cite{Nye2020}, where there is an equal chance to change or remain within an orthant. The objective in their work was to construct a Brownian motion on $\mathcal T^I_{p+1}$ as the limit of a geodesic random walk, and as such may lead to a slower mixing of a Markov chain to sample from the posterior owing to the rather large probability of not changing topology within an update; see Section S2.2 of the Supplementary Material for evidence of this. 

With the prior $\pi$ as the beta-splitting prior for resolved trees or the Poisson-Dirichlet for resolved and unresolved trees, the ability to make local geometry-driven moves on $\mathcal U_p$ leads to a geometric Metropolis-Hastings algorithm that explores $\mathcal U_p$ via the image of an exploration of $\mathcal T_{p+1}$ under $\Psi$, where two acceptance probabilities, both under the reparameterized Bayesian model \eqref{eq:normModel}, corresponding to the decoupled topology and edge length updates are computed to first update the topology and then the edge lengths. Denote $\bX$ as the observed data. The acceptance probability is
\begin{align}\label{eq:MH_AccptRate}
    \max\left\{1, \frac{\pi(T') N_p(\bX; \bm 0,\Psi(T'))q(T \mid T')}{\pi(T) N_p(\bX; \bm 0,\Psi(T))q(T' \mid T)}\right\},
\end{align}
where $q(\cdot \mid \cdot)$ is a transition kernel on $\mathcal T_{p+1}$, applies to updates of both $\cE_{T}$ and $\cL_T$ with suitable changes to $\pi$ and $q$. When updating  $\cE_{T}$ for a fixed $\cL_T$, the transition kernel $q$ is symmetric since because of the uniform probabilities assigned to first choosing an an internal edge to shrink to zero, and second in either choosing a new orthant to move to from the codimension one boundary when sampling resolved trees, or choosing to stay at the orthant boundary or move to an adjacent orthant of codimension at least one when sampling unresolved trees. This implies that \eqref{eq:MH_AccptRate} reduces to
\begin{align}\label{eq:Ratio_Edge}
  \alpha_\cE:=\max\left\{1, \frac{\pi_\cE(\cE_T') N_p(\bX; 0,\Psi(T'))}{\pi_\cE(\cE_T) N_p(\bX; 0,\Psi(T))}\right\}.
\end{align}
where $\pi_\cE(\cE_T)$ is the density from beta-splitting for resolved trees and the Poisson-Dirichler prior for unresolved trees. For updating an edge length $|e_A|\in\cL_T$, for a fixed $\cE_T$, evidently, 
\begin{align}\label{eq:Ratio_Len}
	\alpha_\cL:=\max\left\{1, \frac{ \textrm{Exp}(|e'_A|; a) N_p(\bX; \bm 0,\Psi(T'))\textrm{TN}_{(0,\infty)}(|e_A|; |e'_A|, \sigma_\cL)}{\textrm{Exp}(|e_A|; a) N_p(\bX; \bm 0,\Psi(T))\textrm{TN}_{(0,\infty)}(|e'_A|; |e_A|, \sigma_\cL)}\right\},  
\end{align}
where $\textrm{Exp}(\cdot;a)$ is the exponential density with mean parameter $a$, and the transition kernel $q$ is taken to be $\textrm{TN}_{(0,\infty)}(\cdot ; \mu, \sigma_\cL)$, the truncated normal density on $(0,\infty)$ with mean $\mu$ and standard deviation $\sigma_\cL$. 
The Metropolis-Hastings algorithm for the resolved trees is summarized in Algorithm \ref{algo_MH}. 

\begin{algorithm}[htb!]
\hrulefill
	\small 
	\caption{ Metropolis-Hastings algorithm for posterior sampling}
	\label{algo_MH}
	\SetKwInOut{Input}{Input}
	\SetKwInOut{Output}{Output}
	\SetKwFunction{proc}{}
			\texttt{
	\Input{\vspace*{-5pt}
		\begin{itemize}
			\itemsep -0.5em
			\item[(a)] Initialization: Current iterate $\Sigma^{T_{(l)}}$;
			\item[(b)] Beta-splitting prior $\pi_{\cE}$  and prior $\pi_\cL$ on edge lengths;
			\item[(c)] Number of iterations $M$ and standard deviation $\sigma_\cL$.
		\end{itemize}%
	}
	\Output{Samples from posterior distribution $\nu(\Sigma^T|\bX_1,\ldots,\bX_n)$.}
	\For{$l=1$ \KwTo $M$}{
		\SetKwProg{edgeSet}{Update topology}{}{}
		\edgeSet{}{
			\nl Compute $\Phi(\Sigma^{T_{(l)}})=(\cE_{T_{(l)}},  \cL_{T_{(l)}})$\;
			\nl Randomly remove a split $e_A \in \cE_{T^{(l)}}^I$\;
			\nl  Randomly choose an internal split, say $e_B$, between two splits $\{e_{A'}$, $e_{A''}\}$ different from $e_A$ that are compatible with $\cE_{T_{(l)}}^I\setminus \{e_A\}$ (Proposition \ref{prop1} ensures that there exist two and only two such splits by setting $k=1$ with one internal edge $e_A$ shrunk to zero)\;
			\nl Set edge length $\vert e_B\rvert=\lvert e_A \rvert$\;
			\nl Compute acceptance rate $\alpha_\cE$ from \eqref{eq:Ratio_Edge} and  generate $u\sim \textrm{Unif}(0,1)$\;
			\uIf{$u\leq\alpha_\cE$}{
				\nl Return $\cE_{T_{(l+1)}}=\{e_B\}\cup \cE_{T_{(l)}}\setminus \{e_A\}$\;}
			\Else{
				\nl Return the edge set $\cE_{T_{(l+1)}}=\cE_{T_{(l)}}$\;}
		}
		\SetKwProg{BrLen}{Update edge lengths}{}{}
		\BrLen{}{
			\For{$e_A\in \cE_{T_{(l+1)}}$}{
				\nl Generate the new edge length $|e'_A|$ from $TN_{(0,\infty)}(e_A,\sigma_\cL)$\;
                \nl Compute acceptance rate $\alpha_\cL$ from \eqref{eq:Ratio_Len} and generate $u\sim \textrm{Unif}(0,1)$\;
                \uIf{$u\leq\alpha_\cL$}{
				    \nl Update $\lvert e_A \rvert$=$\lvert e'_A \rvert$\ and return $\cL_{T_{(l+1)}}=\{\lvert e'_A\rvert \}\cup \cL_{T_{(l)}}\setminus \{\lvert e_A\rvert \}$;}
			    \Else{
				    \nl Return $\cL_{T_{(l+1)}}=\cL_{T_{(l)}}$\;}
            }
		}
		\nl Return $\Sigma^{T_{(l+1)}}=\Psi(T_{(l+1)})$.
	}
}
\hrulefill
\end{algorithm}
 For sampling on all of $\mathcal U_p$ when using the Poisson-Dirichlet prior on $\mathcal T_{p+1}$ with mass on unresolved trees, replace the steps 3, 4, and 6 of Algorithm \ref{algo_MH} with 
\begin{itemize}
    \item[\texttt{$3'$}]  \texttt{Randomly choose to select an internal split, say $e_B$, among $(2k+1)!!-1$ splits different from $e_A$ that are compatible with $\cE_{T_{(l)}}^I\setminus \{e_A\}$ or stay at the boundary;}
	\item[\texttt{$4'$}] \texttt{Set edge length $\vert e_B\rvert=\lvert e_A\rvert$ if a new $e_B$ is chosen; otherwise, $\vert e_B\rvert=0$;}
    \item[\texttt{$6'$}] \texttt{Return $\cE_{T_{(l+1)}}=\{e_B\}\cup \cE_{T_{(l)}}\setminus \{e_A\}$ if a new $e_B$ is chosen; otherwise, return $\cE_{T_{(l+1)}}= \cE_{T_{(l)}}\setminus \{e_A\}$.}
\end{itemize}

\subsection{Posterior summaries}\label{sec:postSum}
Algorithm \ref{algo_MH} outputs posterior samples of $\Sigma^{T_{(l)}}, l=1,\ldots,M$, from the posterior distribution $\nu(\Sigma^T|\bX_1,\ldots,\bX_n)$. We consider estimates of two functionals of $\nu(\Sigma^T|\bX_1,\ldots,\bX_n)$: (i) the maximum aposteriori (MAP) ultrametric matrix; (ii) Fr\'{e}chet mean of $\nu(\Sigma^T|\bX_1,\ldots,\bX_n)$ defined as a minimizer of
\begin{equation}
\label{eq:mean}
 \Psi(\tilde T)\mapsto \int d^2(\Psi(T), \Psi(\tilde T))\thinspace \text{d}\mu (\Psi(T)\vert \bX_1,\ldots,\bX_n)\thickspace
\end{equation}
under the reparameterized model \eqref{eq:normModel}. Existence of the Fr\'{e}chet mean minimising the above functional depends on the support of the posterior distribution; when it exists, it will be unique owing to the inherited CAT(0) geometry of $\mathcal U_p$. On $\mathcal T_{p+1}$, the sample Fr\'{e}chet mean $\tilde T$,  obtained by minimising
\[
\mathcal T_{p+1} \ni \tilde T \mapsto \sum_{l=1}^M d_{\text{tree}}^2(\tilde T, T_{(l)})
\]
with respect to the empirical measure on trees $\{T_{(l)}\}$, exists and is unique owing to the CAT(0) geometry of $\mathcal T_{p+1}$ \citep{BH}. From Theorem \ref{thm1}, we thus note that $\Psi(\tilde T)$ estimates the Fr\'{e}chet mean of the posterior $\nu(\Sigma^T|\bX_1,\ldots,\bX_n)$, when its exists. On $\mathcal T_{p+1}$, $\tilde T$ is a consistent estimator of the Fr\'{e}chet mean of $\mu(T|\bX_1,\ldots,\bX_n)$ \citep{BLO}, and by continuity of $\Psi$, $\Psi(\tilde T)$ consistently estimates the Fr\'{e}chet mean of $\nu(\Sigma^T|\bX_1,\ldots,\bX_n)$. 

The estimate $\Psi(\tilde T)$ is computed using the algorithm by \cite{FrechetMeanTr}, and accuracy of approximations is ascertained chiefly with respect to the intrinsic distance $d$ using the polynomial-time algorithm of \cite{BHV_Dist} to compute $d_{\text{tree}}$. Using these tools, we summarize the posterior by quantifying the proportion of subtrees present in $\Sigma^{T_{(l)}}, l=1,\ldots,M$, and construct 95\% credible intervals for each element of the true data generating ultrametric matrix. 
\section{Simulations}\label{sec:Simulation}
We empirically demonstrate the utility of the proposed method through a series of simulation studies and show that the proposed method can restore the underlying ultrametric matrix under different true data generating mechanisms. 

Results for the HMC algorithm is available in the Supplementary Material Section S2.3. For the MH algorithm, we implement Algorithm \ref{algo_MH} for both the binary trees and general multifurcating trees; specifically, we implement (i) the \emph{binary algorithm} with steps 3,4, and 6 and a beta-splitting prior; and (ii) \emph{multifurcating algorithm} with a Poisson-Dirichlet prior with the modified steps $3',4'$ and $6'$ . 

\paragraph{Data generating mechanism.} We generate a tree structure of $p$ leaves as the true underlying tree $T^0$ and map the true tree structure to an ultrametric matrix of dimension $p$. We first consider a resolved tree of $T^0$ and use the binary algorithm to showcase the utility of the proposed method. Given the true ultrametric matrix $\Sigma^{T^0}$, we consider three data generating mechanisms of (i) correct specified normal distribution $X_i \overset{i.i.d.}{\sim} N(0,\Sigma^{T^0})$ and (ii) mis-specified t distribution $X_i \overset{i.i.d.}{\sim} t_{\textrm{df}}(0,\Sigma^{T^0})$ with degrees of freedom four and three ($\textrm{df}=3$ and $4$). We generate the data with five different sample sizes of $n\in\{3p, 5p, 10p, 25p, 50p\}$ and $50$ independent replicates. In this simulation, we obtain the true tree by the function \texttt{rtree} from the \texttt{R} package of \texttt{ape} \citep{EmmanuelParadis2012} and map the tree to the true ultrametric matrix (see Panels (A) and (B) of Figure \ref{fig:EleWise_Covg}). For the unresolved tree, we first generate a resolved tree with $p-2$ internal edges and randomly remove three internal edges. The true unresolved tree and the corresponding ultrametric matrix are shown in Panels (A) and (D) of Figure \ref{fig:multiTr}. Given the corresponding ultrametric matrix, we generate the data with the correctly specified normal distribution of (i) with $n=25p$. We consider the dimension of $p=10$ due to the intensive computation requirement for the competing method. (See Supplementary Material Section S2.7.)


We summarize the posterior samples by using the statistics in Section \ref{sec:postSum} in two aspects: (i) uncertainty quantification via the frequency of splits and element-wise $95\%$ credible interval and (ii) point estimation with the representative matrices. For the \ul{uncertainty quantification}, we first focus on the topology of the matrix of $\cE_T$ and measure the topology recovery in terms of the splits frequency. For each split in posterior matrices, we compute the frequency of the tree topologies in the posterior samples that contain the true splits. We also investigate the coverage for each element in the matrix for the element-wise $95\%$ credible interval. To our best knowledge, no existing method can directly quantify the uncertainty and be considered as the competing method. For the \ul{point estimation}, we calculate the MAP tree and the mean tree \citep{FrechetMeanTr} as representative trees, map the representative trees via $\Psi(\bar{T})$, and measure the matrix norm and the distance $d$ of \eqref{eq:matDist} between the true underlying matrix and the estimated matrices from representative trees. For the binary case, we use the algorithm that focuses on the binary trees and compare the estimated matrices from our method to \citet{MIP_Tr}, which formulates the matrix estimation as a mixed-integer programming (MIP) problem. For the unresolved tree, we use the modified algorithm with a positive mass to stop at the boundary and compare it to the algorithm that considers only binary trees.
Under the matrix norm, we also consider the sample covariance which does not preserve the ultrametric property. 
We assign a beta-splitting prior of $\beta=-1.5$ as the uniform prior on all topology for the binary algorithm and a Poisson-Dirichlet prior of $(\theta,\alpha)=(1,0)$ for the multifurcating algorithm. We consider independent $\textrm{exp}(1)$ on the edge lengths.
We run the MCMC in Algorithm \ref{algo_MH} for $10,000$ iterations and discard the first $9,000$ iterations.

\subsection{Resolved trees}
\begin{figure}[htb!]
    \centering
    \includegraphics[width=\textwidth]{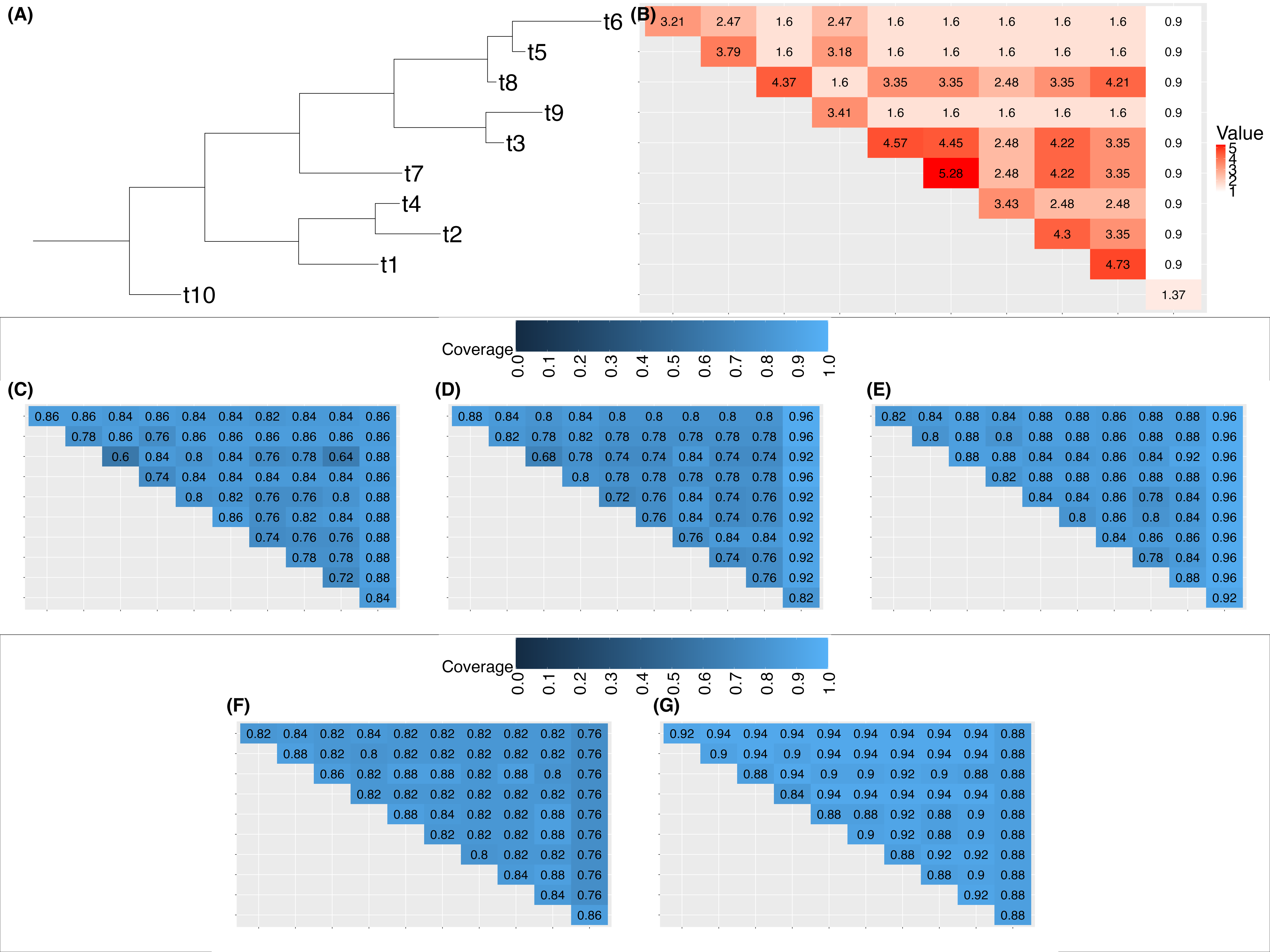}
    \caption{Panels (A) and (B): the true resolved tree and the corresponding ultrametric matrix; (C) to (G): element-wise coverage rates of the nominal $95\%$ credible intervals for the correctly specified normal distribution with five different sample sizes of $30, 50, 100, 250$, and $500$. 
    }
    \label{fig:EleWise_Covg}
\end{figure}

\paragraph{Uncertainty quantification.} We quantify the uncertainty for ultrametric matrices by examining split-wise recovery in Table S1 of Supplementary Material Section S2.4 and element-wise coverage in Figure \ref{fig:EleWise_Covg}. In Table S1, true splits with the corresponding edge lengths from the underlying true matrix are listed in the first row of Table S1. For each true split, we calculate the proportion of the posterior samples that contain the true split shown by each column. The split-wise recovery performs better when the sample size increases for all data generating mechanisms. For different data generating mechanisms, the correct specified model performs the best with sample size of $n=50$ to ensure around $90\%$ of the posterior samples having correct splits. On the other hand, to achieve a similar level of $90\%$ split recovery, the mis-specified t-distribution requires sample size over $100$ and $250$ for $t_4$ and $t_3$, respectively. Among all splits, we also observe that the split with a smaller length ($\lvert e_{5,6}\rvert=0.231$) 
has the worst recovery (mean recovery of $\lvert e_{5,6}\rvert$ is $67.8\%$
while the recovery for other split are around $90\%$ for correct specified model with $n=50$). We present the results of element-wise coverage of the $95\%$ credible interval for the normal distribution in Panels (C) to (G) in Figure \ref{fig:EleWise_Covg}. The results for $t$-distribution are available in Supplementary Material Section S2. Overall, the estimated coverage are high but slightly lower than the nominal coverage (around $0.75$ to $0.94$) and is higher when the sample size increase (medians of the coverage rates for sample size of $(30, 50, 100, 250, 500)$ are $(0.84, 0.78, 0.88, 0.82, 0.90)$).

\begin{figure}[!htb]
    \centering
    \includegraphics[width=\textwidth]{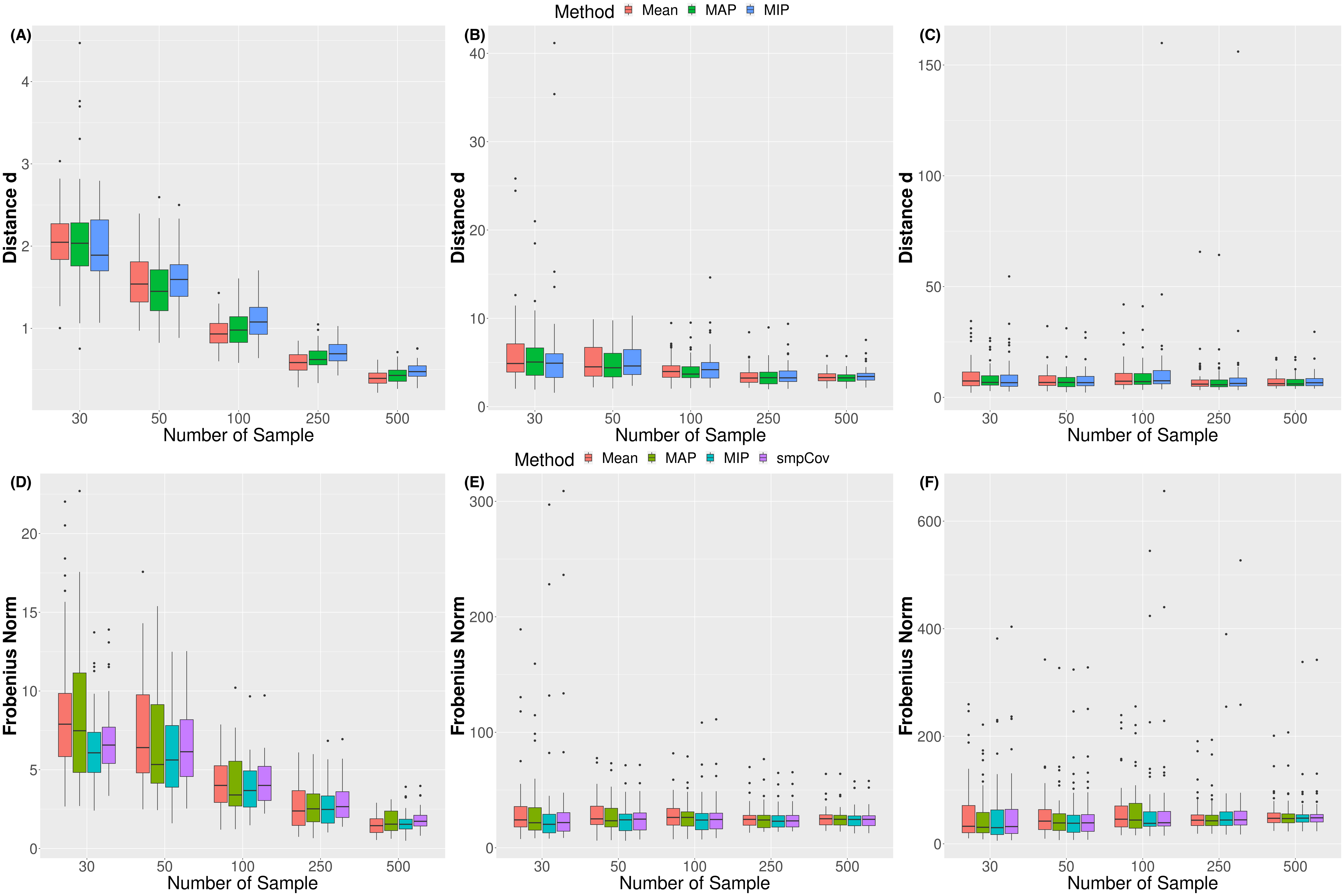}
    \caption{Distances between the estimated matrix and the true matrix under different data generating mechanism and sample sizes. Panels (A) and (D) show results for data generated from the correctly specified normal distribution; (B) and (E) use data from a mis-specified $t_4$ distribution; and (C) and (F) use data from a mis-specified $t_3$ distribution. The mean (red) and MAP (green) from our method is comparable to competing methods (blue for MIP and purple for sample covariance) in terms of the distance $d$ from \eqref{eq:matDist} ((A) to (C)) and matrix norm ((D) to (F)).}
    \label{fig:SimDist}
\end{figure}

\paragraph{Point estimation.}
We show the distance from the estimated matrices to the true matrix in Figure \ref{fig:SimDist}. We observe that the matrices mapped from the mean and MAP trees from our method are comparable to the estimated matrix from \citet{MIP_Tr} and sample covariance in terms of matrix norm (Panel (A) to (C)) and distance $d$ from \eqref{eq:matDist} (Panel (D) to (F)) across different data generating mechanisms and sample sizes. When the model is correctly specified (Panel (A) and (D)), all methods benefit from the increase of the sample size with a smaller distance to the true matrix (medians of matrix norm and distance $d$ for sample sizes of $(30, 50, 100, 250, 500)$ are $(6.75, 5.89, 3.91, 2.50, 1.56)$ and $(2.01, 1.53, 0.987, 0.627, 0.435)$, respectively). For the mis-specified scenario (Panel (B) and (E) for $t_4$, and (C) and (F) for $t_3$), the advantage from the larger sample size is moderate (medians of matrix norm and distance $d$ for sample sizes of $(30, 50, 100, 250, 500)$ are $(21.9, 24.6, 25.2, 23.7, 24.8)$ and $(4.97, 4.46, 3.97, 3.25, 3.28)$ for $t_4$, and $(31.6, 39.5, 42.5, 44.0, 47.5)$ and $(7.00, 6.70, 7.39, 6.10, 6.28)$ for $t_3$).

We also provide additional results for simulation including (i) convergence diagnostics, (ii) element-wise coverage for mis-specified t-distribution, (iii) the topology trajectory for the proposed method, (iv) the simulation results for the data generated from a underlying tree with the same sum of the edge lengths from the root the all leaves, and (v) the simulation results using the Hamiltonian Monte Carlo. Comparing to other methods (e.g. MIP), we also observe that our proposed MCMC is relative scalable in terms of the number of leaves. Specifically, when the number of leaves is doubled from $p=10$ to $p=20$, the time required for MIP is $21.5$ times, while it is only $3.5$ times for the proposed method. All additional simulation studies are provided in Supplementary Material Section S2.

\subsection{Unresolved trees}
\paragraph{Uncertainty quantification.}
We compare the proposed method with two versions of multifurcating and binary algorithms. Panels (B) and (C) of Figure \ref{fig:multiTr} illustrate one tree (MAP) drawn from the multifurcating and binary algorithms, respectively. Obviously, the multifurcating algorithm generates the trees with both non-binary and binary nodes, while the binary algorithm sticks to the binary nodes even when the true underlying tree is unresolved. 
For split-wise recovery, we observe that both the binary and multifurcating algorithms perfectly recover five splits, with $100\%$ of the posterior samples from both algorithms including all five. Moreover, the estimated coverage from both algorithms are comparably high (Panels (E) and (F) of Figure \ref{fig:multiTr}; median of the coverage rates for multifurcating algorithm: $0.88$, and binary algorithm: $0.94$). Although the binary algorithm seemingly recovers the true split, it generates redundant splits (three redundant splits), while the multifurcating algorithm generates much less redundant splits (mean number of splits: $5.27$). 
For example, three additional splits of $e_{1,4,6,10}, e_{1,6,10}$, and $e_{1,3,4,5,6,8,9,10}$ are included in the tree generated from the binary algorithm in Panel (C), while the tree of Panel (B) from the multifurcating algorithm perfectly aligns with the true topology.


\paragraph{Point estimation.}
The distance between the estimated matrices and the true matrix is shown in Panels (G) and (H) of Figure \ref{fig:multiTr}. The point estimators of the mean and MAP matrix generated from the multifurcating algorithm slightly outperform those of the binary algorithm in the distance $d$ (median of MAP and mean for multifurcating: $(0.411, 0.514)$ and binary: $(0.445, 0.552)$). For the Frobenius norm, the mean matrix of the multifurcating algorithm (median: $1.07$) is slightly better than the one from the binary algorithm ($1.31$). Moreover, the multifurcating algorithm (median: $1.53$) generates the MAP matrix that is comparable to that of the binary algorithm ($1.45$). Overall, we see that the distance between the true matrix and the estimated matrix of the binary algorithm is only worse than that of the multifurcating by a small margin. We conjecture that the binary algorithm moves close to the true unresolved tree by shrinking the redundant branch lengths to small values. For example, three additional splits of the MAP tree from the binary algorithm in Panel (B) are relatively small ($\lvert e_{1,4,6,10}\rvert=0.194, \lvert e_{1,6,10}\rvert=0.059$, and $\lvert e_{1,3,4,5,6,8,9,10}\rvert=0.180$ with the rest of branch lengths $>0.41$). 

The proposed method of both the binary and multifurcating algorithms generate posterior samples that recover the underlying matrix well. 
While the multifurcating algorithm draws posterior samples that are closer to the true unresolved tree by allowing the multifurcating nodes, the binary algorithm recovers the underlying topology and shrinks the redundant splits to smaller branch lengths. Meanwhile, generating multifurcating nodes results in a heavier computational requirement due to the larger $(2k+1)!!$ candidate orthants. 

\begin{figure}[!htb]
    \centering
    \includegraphics[width=\linewidth]{Figures/multiRes.png}
    \caption{Panels(A) and (D): underlying multifurcating tree and the corresponding ultrametric matrix. (B) and (C): Point estimation of the MAP tree drawn from the multifurcating and binary algorithms on one random replicate. (E) and (F): element-wise coverage rates of the nominal $95\%$ credible interval generated from the multifurcating and binary algorithms over $50$ replicates. (G) and (H): the distance $d$ and Frobenius norm between the true matrix and the estimated matrix over $50$ replicates.}
    \label{fig:multiTr}
\end{figure}

\ul{In summary}, the geometry of the set $\cU_p$ provides four main inferential and computational advantages comparing to existing models: (i) enables the uncertainty quantification on ultrametric matrices along with the point estimator, (ii) is robust to model mis-specifications, (iii) incorporates the multifurcating tree naturally, and (iv) is computational efficient in terms of convergence and scalability of number of leaves.

\section{Analysis of Treatment Tree in Cancer}\label{sec:DataAnalysis}

We illustrate our proposed method on a pre-clinical patient-derived xenograft (PDX) data to discover promising cancer treatments. Due to the impracticality of testing multiple treatments on the same patient in actual clinical trials, PDXs involve an experiment design that evaluates multiple treatments administered to samples obtained from human tumors implanted into genetically identical mice. The mice are then treated as the ``avatars'' to mimic potential responses to different treatments. We leverage a PDX dataset of \ul{N}ovartis \ul{I}nstitutes for \ul{B}ioMedical \ul{R}esearch - \ul{PDX} \ul{E}ncyclopedia [NIBR-PDXE, \citep{pmid26479923}] that has collected over $1,000$ PDX lines across multiple cancers with a $1 \times 1 \times 1$ design (one animal per PDX model per treatment). 

For our analysis, we focus on cutaneous melanoma (a type of skin cancer) which consists of $14$ treatments and $32$ PDX lines. The primary response is the tumor size difference before and after treatment administration, following the approach by \citet{doi:10.1080/01621459.2020.1828091}, with the untreated group as the reference group. Positive responses indicate that the treatment shrunk the tumor more than the untreated group with a higher value representing a better efficacy. We posit that treatments with similar mechanism should induce similar levels of responses, and we aim to construct a hierarchical tree structure to infer the mechanistic similarity based on the treatment responses. We use the binary algorithm since scientifically, the assumption of multifurcating nodes implies that all pairs of treatment in the subtree attached to the multifurcating node share an equal mechanistic similarity, which is unrealistic due to the cancer heterogeneity \citep{CanHet_Jack2018} and unexpected off-target effects in the pre-clinical trial \citep{offTarget_Lin2019}. 
We ran Algorithm \ref{algo_MH} with a beta-splitting prior of $10,000$ iterations and discarded the first $9,000$ iterations. 
The convergence diagnostics of the algorithm are shown in Supplementary Material Section S3. Herein, we summarize the results with the MAP and mean trees and highlight subtrees with frequency above $90\%$.

\begin{figure}
    \centering
    \includegraphics[width=\textwidth]{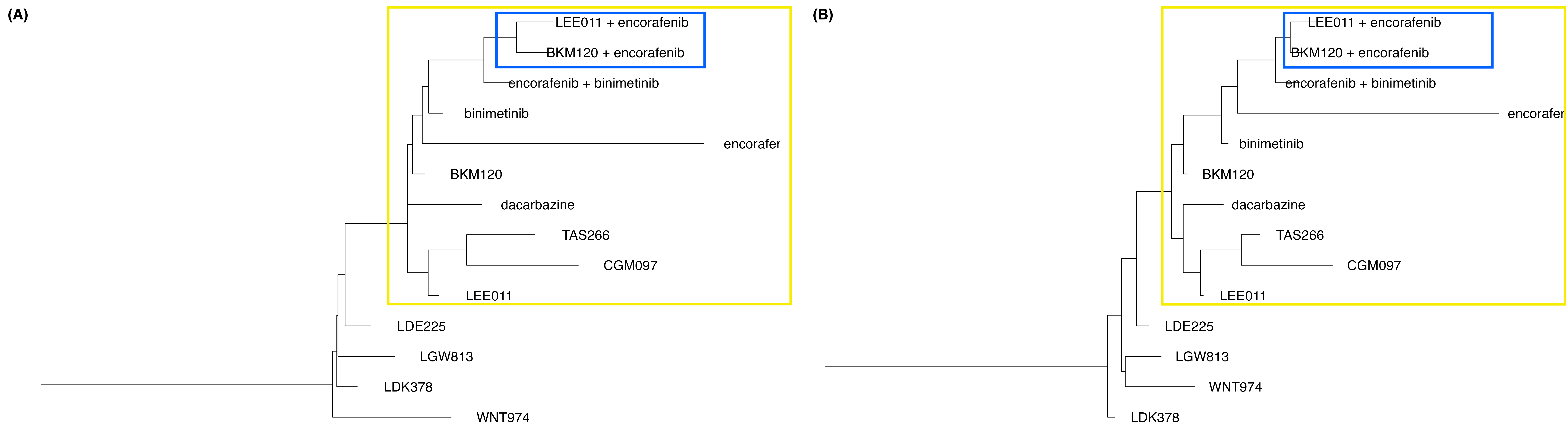}
    \caption{The mean (Panel (A)) and MAP trees (Panel (B)) for the melanoma. Two boxes emphasize the subtrees with high frequencies ($>90\%$) in the posterior samples: blue: $91\%$, and yellow: $98\%$.}
    \label{fig:TrtTr}
\end{figure}

Figure \ref{fig:TrtTr} shows the mean (Panel (A)) and MAP trees (Panel (B)) with subtrees that consistently appear in the posterior samples. Two subtrees with frequencies higher than $90\%$ are emphasized by boxes: blue ($91\%$), and yellow ($98\%$). We observe that the MAP and mean trees share many subtrees with the same topology. For example, the subtrees in the boxes are identical in both the mean and MAP trees. Additionally, two combination treatments (highlighted by blue surrounding box) form a tight sub-tree that appears in over $90\%$ of posterior samples, indicating a high level of mechanistic similarity of the combination therapies. Two combination therapies consist of two agents, with one agent being encorafenib and the other targeting one of the following pathways: phosphoinositide 3-kinases (BKM120), and cyclin-dependent kinases (LEE011). As these pathways are closely related and share common downstream mechanisms \citep[e.g.,][]{pmid29547722, pmid30974877}, it is not surprising to see that all combination therapies form a tight subtree in the tree structure.

\section{Discussion}\label{sec:Discussion}

There are several future directions that may further improve the utility of the proposed method. First, in certain contexts, latent tree models for phylogenetic studies assume that trees have edge lengths proportional to the length of time between speciation events represented by each node. This results in a sum constraint on the edge lengths from the root to leaves. Geometry of the BHV treespace is unsuitable for such trees. Suitable modification \citep[e.g.,][]{GD} that results in a similar stratified space, to which the current geometric framework can be extended, can be used. 

Second, the bijection $\Phi:\mathcal U_p \to \mathcal T_{p+1}$, and the algorithm for its computation, may lead to a useful set of coordinates in terms of the basis matrices $\{E_A\}$ and coefficients $\{|e_A|\}$ for further study of $(\mathcal T_{p+1}, d_{\text{tree}})$, especially in simplifying computations presently used on the treespace. 

Third, from a modelling perspective, for trees containing a large number of leaves (large $p$), the prior construction in Section \ref{sec:TrPrior} makes possible development of Bayesian models based on principled variable selection priors that decouples topology and edge length information on latent trees. For example, spike-and-slab priors on the coefficients in the representation \eqref{eq:SigmaT_Decomp} will allow modelling of sparse representations of $\Sigma^T$ corresponding to trees $T$ of a fixed topology.

Fourth, the permutation group on $\{0,1\ldots,p\}$ is the isometry group of $\cT_{p+1}$ \citep{grindstaff2020isometry}; this implies that $d(P\Sigma^T_1P^\transp,P\Sigma^T_2P^\transp)=d(\Sigma_1^T,\Sigma_2^T)$ for all $p \times p$ permutation matrices $P$. Moreover, since $\cU_p$ is closed under simultaneous permutations of its rows and columns \citep{McCullagh2006StructuredCM}, sampling on $\cU_p$ with updates $P\Sigma^TP^\transp$ that change the topology of $T$ via a judicious prior on permutation matrices $P$ may lead to better exploration of $\cU_p$ that is compatible with its intrinsic geometry. 


Finally, a natural extension to the latent tree model is to introduce covariates $\bm z$ through the decomposition \eqref{eq:SigmaT_Decomp} that leads to case-specific ultrametric covariance matrices $\Sigma^{T(\bm z)}$. A prior distribution using the construction in Section \ref{sec:TrPrior} will assign weights on different tree topologies and edge lengths depending on $\bm z$, while its fragmentation process-based definition will allow borrowing of information as $\bm z$ varies. We leaves these topics for future research. 

\section*{Acknowledgements}
KB acknowledges support from grants  NSF 2015374 and NIH R37-CA21495. The work is partly supported by a Michigan Institute for Data Science (MIDAS) seed grant to ZW. VB acknowledges support from grants NIH P30 CA046592 and NIH R01 CA244845-01A1.

\section*{Appendix}
The following Lemma proves positive definiteness of the ultrametric matrix corresponding to any tree $T \in \mathcal T_{p+1}$, including those on the boundary. 
\begin{lemma}
\label{lem1}
$\Sigma^T$ corresponding to any tree $T \in \mathcal T_{p+1}$ is a positive definite strictly ultrametric matrix. 
\end{lemma}
\begin{proof}
The subset $\{\bm 0\} \times \mathcal X_{p+1}$ of $\mathcal T_{p+1}$ corresponds to the set of star trees with a nonnegative root edge, i.e. the set of trees with internal edges of length zero, $p$ leaf edges of positive length, and a root edge of nonnegative length. Theorem 2.2 of \cite{Nabben1994} describes how such trees map to matrices of the form
\[
d_1 \bm 1 \bm 1^\transp + \sum_{i=1}^p d_{j_i}\bm u_i \bm u_i^\transp,
\]
where $\{\bm u_i, i=1,\ldots,p\}$ is the standard basis vectors in $\mathbb R^p$ with one in the $i$th position and zero elsewhere, and $d_{j_1},\ldots, d_{j_p}$ is a subsequence from $\{d_i\}$ in \eqref{eq:SigmaT_NaiveDecomp} corresponding to leaf edges, and $d_1 \geq 0$ is the length of the root edge. Clearly, such matrices satisfy property (ii) in Definition 1, and are strictly ultrametric. They are also positive definite. To see this, note that the second term in the sum gives rise to positive definite diagonal matrices with positive entries. For such a diagonal matrix $W$, note that $W+d_1 \bm 1 \bm 1^\transp$ is its rank-one perturbation with scale $d_1>0$, symmetric, diagonally dominant, and hence invertible and positive definite. 

Growing an internal edge to a tree at the origin to move to the boundary component of codimension $(p-1)$ translates to the operation
\[
d_1 \bm 1 \bm 1^\transp + \sum_{i=1}^p d_{j_i}\bm u_i \bm u_i^\transp+d_{j_{p+1}} \bm v \bm v^\transp, 
\]
where $\bm v$ is a binary vector and $\bm v \bm v^\transp$ is of rank 1 with $d_{j_{p+1}}>0$, on ultrametric matrices. Using the argument from above, such matrices are positive definite and strictly ultrametric. Adding additional internal edges amounts to performing repeated rank-one perturbations of a positive definite matrix with positive $d_{j_k}, k=p+2,\ldots,2p-1$, which preserves strict ultrametricity and positive definiteness, which a zero-valued $d_{j_k}$ leaves the property unaffected.
\end{proof}
\subsection*{Proof of Theorem \ref{thm1}}
\begin{proof}
We first show that $\Phi$ is surjective. Consider a tree $T$ in $\mathcal T_{p+1}$, which may be resolved or unresolved. For each leaf edge $e_{A}$ for a singleton split $A=\{i\}$ with $i \in [p]$, let $v_{i+1}$ be a vector in $\mathbb R^p$ with 1 in the $i$th position and zero everywhere else; when $A=\{0\}$ let $v_1$ be the vector of ones in $\mathbb R^p$. There are thus $p+1$ vectors corresponding to the leaves of $T$. When $e_A, A \subset L$ is an internal edge, let $v_j \in \mathbb R^p$ for $j=p+2,\ldots,K$ contain ones in component indices in $A$ and zero elsewhere, where $K \leq 2p-1$, with equality for resolved trees. Let $V=(v_0,v_1,\ldots,v_{K}) \in \mathbb R^{p \times K}$, and $D \in \mathbb R^{K \times K}$ be the diagonal matrix of edge lengths $|e_{A}|, A \subset L$.

From \eqref{eq:SigmaT_Decomp}, the matrix $VDV^\transp$ will be an ultrametric matrix if $V$ has the partition property: for every column $v_i$ in $V$ there exists two distinct columns $v_j$ and $v_k$ such that $v_i=v_j+v_k$. For each edge $e_A$ associated with a split $A$ define the unique $p$-dimensional binary vector $b_A$ with ones at indices that are in $A$ and zero for indices in $A^c$. Arrange the vectors into a $p \times K$ binary matrix $B=(b_{A_1},\ldots,b_{A_{K}})^\transp$. In order to relate the columns of $B$ to those of $V$ possessing the partition property, we use the logical \texttt{AND} operator $\wedge$ on columns of $B$. In other words, the compatibility criterion that one of $A_1\cap A_2, A_1\cap A_2^c, A_1^c\cap A_2$ associated with two splits $e_{A_1}$ and $e_{A_2}$ be empty translates to one of $b_{A_1} \wedge b_{A_2}, \bar b_{A_1} \wedge b_{A_2}, b_{A_1} \wedge \bar b_{A_2}$ equalling the zero vector $\bm 0$, where $\bar b$ is the negation of $b$. 

Denote by $\vee$ the logical \texttt{OR} operator. Note that $v_i=v_j+v_k$ if and only if $b_{A_i}=b_{A_j} \vee b_{A_k}$ and $b_{A_j} \wedge b_{A_k}=\bm 0$ while each of $\bar b_{A_j} \wedge b_{A_k}, b_{A_j} \wedge \bar b_{A_k}$ does not equal $\bm 0$, rendering the splits corresponding to the pair $(b_{A_j}, b_{A_k})$ compatible. Similarly, $b_{A_i} \wedge b_{A_j}\neq \bm 0$, and only one of $\bar b_{A_i} \wedge b_{A_j}, b_{A_i} \wedge \bar b_{A_j}$ equals $\bm 0$, since otherwise $b_{A_i} \neq b_{A_j} \vee b_{A_k}$. The splits corresponding to pair $(b_{A_i}, b_{A_j})$ are hence compatible; a similar argument is used to show compatibility of $(b_{A_i}, b_{A_k})$. The map $\Phi$ is hence surjective. 

Injectivity will be established by constructing a unique tree $T$ in $\cT_{p+1}$ for given ultrametric matrix $\Sigma^T$. For any $K \times K$ permutation matrix $\Pi$ observe that $\Pi D \Pi^\transp$ fixes the diagonal $D$. Consequently, the transformation $V \mapsto V\Pi$ leaves $V\Pi D \Pi ^\transp V^\transp$ invariant, but permutes the columns of $V$. Moreover, given an ultrametric matrix $\Sigma^T$, the value of $K$ that determines the codimension of the stratum in $\mathcal T_{p+1}$ to which the tree $T$ belongs is not directly available.  

Let $\tilde \Sigma^T=\Sigma^T-\alpha_1\bm 1 \bm 1^\transp$, where $\alpha_1 \geq 0$ is the smallest element of $\Sigma^T$, and corresponds to the root edge of $T$, whose vertex representation is $(0,\text{root})$. Then, the pattern of zeros of $\tilde \Sigma$ characterizes the number of children of the root. In particular, the $k_1$ columns of $\tilde \Sigma^T$ with unique pattern of zeros implies that the root has $k_1$ children; if $k_1>2$, then $T$ is unresolved. A straightforward modification of the proof of Proposition 2.1 of \cite{Nabben1994} to accommodate the possibility that $k_1>2$ ensures existence of a $p \times p$ permutation matrix $P_1$ such that 
\[
P_1\tilde \Sigma^T P_1^\transp=
\begin{pmatrix}
C_1&0&\cdots&0\\
0&C_2&\cdots &0\\
\vdots & 0&\ddots & 0\\
0&0 &\cdots &C_{k_1}
\end{pmatrix},
\] 
where $C_1,\ldots,C_{k_1}$ are ultrametric matrices. The conjugate action of $P_1$ via simultaneous permutation of the rows and columns of $\tilde \Sigma^T$ ensures that edge relationships between the nodes of $T$ are preserved. 
At this stage we have uncovered a portion of the tree consisting of the root edge of length $\alpha_1 \geq 0$ connecting to root with $k_1$ children. A choice of labeling of the children amounts to an ordering of the blocks $C_i$ along the diagonal within $\tilde \Sigma^T$. Note, however, that although the permutation $P_1$ is not unique, the subset $A_1 \subset [p]$ of row and column indices of $\tilde \Sigma ^T$ within each $C_i$ is unique owing to the decomposition property in \eqref{eq:SigmaT_Decomp}; their ordering within each $C_i$ may be arbitrary. Thus, relabeling the $C_i$ is a permutation of the children of the root. 

Repeating the procedure above on $C_1$ by determining another $|C_1| \times |C_1|$ permutation matrix $P_2$ such $P_2(C_1-\alpha_2 \bm 1 \bm 1^\transp)P_2^\transp$ is block diagonal containing $k_2$ ultrametric matrices, and determines $k_2$ number of children at node $u_1$ corresponding to the edge $e=(\text{root},u_1)$ of length $\alpha_2$. Continuing in this fashion will uncover the subtree descendant from $u_1$, where at each stage $j$ a choice of ordering of the diagonal blocks of ultrametric matrices will determine the ordering of the children at node $u_j$; the process will eventually result in a diagonal matrix of dimension $m \times m$, where $m \in \{2,\ldots,p\}$ with the blocks being positive scalars ($1 \times 1$ ultrametric matrices), which determines the leftmost subtree of $T$ with root $u_1$ and leaves given by $m$ distinct elements of $L$. The $m$ positive scalars comprise a subset of the $p$ positive coefficients $d_j$ in \eqref{eq:SigmaT_Decomp} and the non-negatives $d_j$s correspond to the $\alpha_j$s. 
This process is repeated on each block $C_j, j=1,\ldots,k_1$ to uncover the descendant subtrees.  

The tree $T$ uncovered in this manner is unique up to permutation of the node labels of the $k_j$ children of the node $u_j$ descendant from the root of the subtree at stage $j$ of the recursive process. The number of columns $K$ of $V$ equals the sum of the $k_j$s obtained from the recursive procedure. Cumulatively, these permutations are encoded in the permutation $\Pi$ on the columns of the aforementioned $V$. The partition property of an ultrametric matrix ensures that the ordering of ultrametric matrices in the block diagonal matrix in each stage of process does not break the combinatorial structure of the descendant subtrees and the edge lengths. In other words, two trees $T_1$ and $T_2$ corresponding to two permutations $\Pi_1$ and $\Pi_2$ of the columns of $V$  differ in only two ways: (i) by a permutation of the labels at the internal nodes; (ii) swapping labels of left and right children; changes dues to (ii) do not change the tree topology, and from definition the BHV treespace, (i) has no bearing on the orthant in which $T$ lies in. Thus, on $\cT_{p+1}$ the trees $T$ lie in the same orthant with the same coordinates coming from their edge lengths, so that the map $\Phi$ is injective.

The BHV space $(\mathcal T_{p+1}^I, d_{\text{BHV}})$ is a CAT(0) space \citep[Lemma 4.1]{Billera2001}. The space $(\mathcal X_{p+1}, \| \cdot \|_2)$ is Euclidean and hence CAT(0), and $(\mathcal T_{p+1}, d_{\text{tree}})$ as a product of two CAT(0) spaces is thus CAT(0) \citep{BH}. With the bijective $\Phi$ we can hence pull back the nonpositive curvature of $\mathcal T_{p+1}$ onto $\mathcal U_p$ making it globally CAT(0).
	
Inclusion of the leaf edge lengths ensures that each stratum of $\mathcal T_{p+1}$ can now be identified with the product $\mathbb R_{\geq 0}^{p-2} \times \mathcal X_{p+1}$, as a stratified space with the strata glued isometrically along their shared boundaries of the first factor in the product.  Then $\mathcal U_p$ is prescribed a stratification under $\Psi$ with the same $(2p-3)!!$ number of strata each of the same dimension as that of $\mathcal T_p$. The proof is now complete. 
\end{proof}

\subsection*{Proof of Proposition \ref{prop1}}
\begin{proof}
	On a codimension one boundary, from the decomposition in \eqref{eq:SigmaT_Decomp}, we note that the unresolved tree corresponding to $\tilde \Sigma ^T$ is obtained by creating a degree 4 vertex. Starting from such an unresolved tree there are exactly three possible ways to grow an edge and create a fully resolved tree. This procedure identifies the three unique ultrametric matrices $\Sigma^T_p, \Sigma^T_q, \Sigma^T_r$ on three distinct strata $p,q,r$. The claim on $k$-codimensional boundaries follows from the corresponding picture in the BHV space \citep{Billera2001}. The order inequality on $\tilde \Sigma^T$ follows upon noting that an element $\Sigma^T_{ij}$ of $\Sigma^T$ is the sum of edge lengths from the root to the most recent common ancestor of $i$ and $j$. 
\end{proof}

\bibliography{UltraMat}
\end{document}


\setstcolor{red}

\def\spacingset#1{\renewcommand{\baselinestretch}%
{#1}\small\normalsize} \spacingset{1}
\input{definitions}


\if1\blind
{
  \title{\bf Supplementary Material for ``Geometry-driven Bayesian Inference for Ultrametric Covariance Matrices''}
  \author{Tsung-Hung Yao$^{a,*}$, Zhenke Wu$^{b}$, Karthik Bharath$^{c}$,\\
  and Veerabhadran Baladandayuthapani$^{b}$\\~\\
  $^{a}$Department of Biostatistics, The University of Texas MD Anderson Cancer Center\\
  $^{b}$Department of Biostatistics, University of Michigan\\
  $^{c}$School of Mathematical Sciences, University of Nottingham\\
  $^{*}${\small corresponding author. E-mail: yaots@umich.edu}
			}
  \maketitle
} \fi

\if0\blind
{
  \bigskip
  \bigskip
  \begin{center}
    {\Large\bf Supplementary Material for ``Geometry-driven Bayesian Inference for Ultrametric Covariance Matrices''}
\end{center}
  \medskip
} \fi

\bigskip

\spacingset{1.45} 

\section{Hamiltonian Monte Carlo}\label{supp:HMC}
Following the Hamiltonian Monte Carlo (HMC) algorithm from \citet{ppHMC_Dinh2017}, we propose an HMC algorithm for ultrametric matrices. Based on Corollary 1, denote the tree-structured covariance of size $p$ by $p$ as $\bSigma^T=\sum_{e_A\in\cE_T} \lvert e_A\rvert E_A$, where $\cE_T$ is the edge set, $\lvert e_A\rvert$ are the positive and continuous branch lengths, and $E_A$ are the binary matrices that jointly represent the tree topology. The core concept of HMC is to improve the acceptance rate of the Metropolis–Hastings algorithm (MH) by proposing an ideal proposal via the classical Hamiltonian equations, which requires the parameters of interest to be continuous. In this case, we first focus on the case within the same orthant with a fixed topology. For the ease of notation, assume $\lvert\cE_T\rvert=Q, p\leq Q\leq 2p-1$ and enumerate the edges $e_A\in\cE_T$ with $d_j=\lvert e_A\rvert$ and the corresponding binary matrices $E_A=\bV_j=\bv_j\bv_j^\transp$ for $j=1,\ldots,Q$. Let $\ba=[a_1,\ldots,a_Q]^\transp$ be the momentum for the branch lengths and $H(\bSigma^T,\ba)=U(\bSigma^T)+K(\ba)$ be the Hamiltonian for the tree-structure covariance as the sum of the position energy, $U(\bSigma^T)$, and the kinetic energy, $K(\ba)$. In this paper, we let $U(\bSigma^T)=-\log p(\bSigma^T\mid \bX)$ be the negative posterior distribution under logarithm and the kinetic energy $K(\ba)=\frac{1}{2}\ba^\transp \bM^{-1} \ba=\sum_{j=1}^{Q} \frac{a_j^2}{2m_j}$ be the kernel of the multivariate Gaussian kernel with diagonal covariance $\bM=\textrm{diag}(m_1,\ldots,m_{Q})$. 

One important step of the Hamiltonian Monte Carlo (HMC) is to propose through the Hamiltonian dynamic $\frac{d}{dt}a_j=-\frac{\partial U({\bf \Sigma}^T)}{\partial d_j}$ and $\frac{d}{dt}d_j= \frac{a_j}{m_j}$. To do so, the gradient of the posterior distribution is obtained by $\frac{\partial U({\bf \Sigma}^T)}{\partial d_j} = -\frac{\partial \log p({\bf \Sigma}^T\mid \bX)}{\partial d_j} = -\frac{\partial}{\partial d_j} \left\{\log p(\bX\mid \bSigma^T) +\log\pi(\bSigma^T) \right\}$. The derivative for the first term of the normal likelihood can be easily obtained by
\begin{align*}
    \frac{\partial \log \bN(\bX;\mathbf{0},\bSigma^T)}{\partial d_j} & = -\frac{1}{2}\frac{\partial}{\partial d_j} \left\{n\log\textrm{det}\left(\sum_{j=1}^{Q}d_j\bv_j\bv_j^\transp\right) + \sum_{i=1}^n\bx_i^\transp \left(\sum_{j=1}^{Q}d_j\bv_j\bv_j^\transp\right)^{-1} \bx_i \right\}.
\end{align*}
Denote $\bSigma_{-j}=\sum_{k\neq j}^{Q}d_k\bv_k\bv_k^\transp$. By using the Woodbury formula and Sylvester's determinant theorem, we obtain 
\begin{align*}
    \left(\bSigma^T\right)^{-1}&=\left(\bSigma_{-j} + d_j\bv_j\bv_j^\transp\right)^{-1} = \bSigma_{-j}^{-1} - \frac{d_j \bSigma_{-j}^{-1}\bv_j\bv_j^\transp \bSigma_{-j}^{-1}}{1+d_j\bv_j^\transp \bSigma_{-j}^{-1} \bv_j}, \thickspace \frac{\partial}{\partial d_j} \left(\bSigma^T\right)^{-1} = - \frac{\bSigma_{-j}^{-1} \bv_j \bv_j^\transp \bSigma_{-j}^{-1}}{(1+d_j\bv_j^\transp \bSigma_{-j}^{-1}\bv_j)^2}, \\
    \textrm{det}(\bSigma^T) &= 
    \textrm{det}(\bSigma_{-j}) \left(1+d_j\bv_j^\transp \bSigma_{-j}^{-1}\bv_j\right), \textrm{ and } \frac{\partial}{\partial d_j} \textrm{det}(\bSigma^T) = \textrm{det}(\bSigma_{-j}) \bv_j^\transp \bSigma_{-j}^{-1}\bv_j.
\end{align*} 
The derivative of the normal likelihood then becomes
\begin{align*}
    -\frac{1}{2}\left\{\frac{n\bv_j^\transp \bSigma_{-j}^{-1}\bv_j}{1+d_j\bv_j^\transp \bSigma_{-j}^{-1}\bv_j} - \sum_{i=1}^n \bx_i^\transp \frac{\bSigma_{-j}^{-1} \bv_j \bv_j^\transp \bSigma_{-j}^{-1}}{(1+d_j\bv_j^\transp \bSigma_{-j}^{-1}\bv_j)^2}\bx_i\right\}.
\end{align*}
With the assumption of the exponential prior on the branch lengths, $d_j\overset{i.i.d.}{\sim}\textrm{Exp}(\lambda)$, the gradient of the potential energy is obtained by
\begin{align*}
    \frac{\partial U({\bf \Sigma}^T)}{\partial d_j} = \frac{1}{2}\left\{\frac{n\bv_j^\transp \bSigma_{-j}^{-1}\bv_j}{1+d_j\bv_j^\transp \bSigma_{-j}^{-1}\bv_j} - \sum_{i=1}^n \bx_i^\transp \frac{\bSigma_{-j}^{-1} \bv_j \bv_j^\transp \bSigma_{-j}^{-1}}{(1+d_j\bv_j^\transp \bSigma_{-j}^{-1}\bv_j)^2}\bx_i\right\} + \lambda.
\end{align*}
Given the step size of $\epsilon$, we can propose new branch lengths by the leapfrog algorithm as follows.
\begin{align}
    \nonumber\bar{\ba}&=\ba - \frac{\epsilon}{2} \nabla_d U({\bf \Sigma}^T)\\
    \label{eq:HMC_updateBr}\bd'&=\bd+\epsilon \bar{\ba}\\
    \nonumber\ba'&= \bar{\ba} - \frac{\epsilon}{2} \nabla_d U({\bf \Sigma}^T)
\end{align}
However, Equation \eqref{eq:HMC_updateBr} does not guarantee the positiveness of the new proposed $\bd'$. The change in sign of the proposed $\bd'$ from positive to negative implies that the algorithm is oriented to the boundary of the current orthant and jumps to a new nearby orthant. 

Now, we generalize the algorithm considering different orthants. When one of the proposed branch lengths changes the sign from positive to negative, the HMC algorithm hits the boundary of the codimension of one. We then randomly choose one of the compatible splits uniformly, flip the sign of the element of the momentum, and match the momentum to the new split. Note that the current split is excluded when choosing a new nearby orthant to improve the algorithm. In practice, multiple elements of new $\bd'$ might cross the boundaries given a step size of $\epsilon$ and a momentum $\ba$. Consider $\epsilon_j^0\in(0,\epsilon]$ such that $d'_j+\epsilon_j^0a_j=0$ and $\epsilon_j^0=\infty$ if $d'_j+\epsilon a_j>0$. Elements with finite $\epsilon_j^0$ cross the boundaries. If we view Equation \eqref{eq:HMC_updateBr} as a path for the algorithm to traverse from $\bd$ to a proposed matrix with a total step size $\epsilon$, each $\epsilon_j^0$ becomes the fractured step size for $d_j$ to reach the boundary. We construct the proposal along the path with a pseudo-time $t_\epsilon\in[0,\epsilon]$. Specifically, consider a path with two fractured step sizes $\epsilon_j^0<\epsilon_{j'}^0<\epsilon$. When the path hits the boundary at $t_\epsilon=\epsilon_j^0$, we flip the sign of $a_j$ and assign $a_j$ to one of the compatible splits. From $t_\epsilon=\epsilon_j^0$, we then move the algorithm by the step size $\epsilon_{j'}^0 - \epsilon_j^0$ to reach the boundary with $d_{j'}=0$ at $t_\epsilon=\epsilon_j^0+\epsilon_{j'}^0 - \epsilon_j^0=\epsilon_{j'}^0$ and select one of the candidate splits with $a_{j'}$ of the flipped sign. Finally, the algorithm proceeds to the final proposal with a step size of $\epsilon-\epsilon_{j'}^0$. For example, consider a matrix with $p=4$, and we focus on two internal edges $(\lvert e_{12}\rvert,\lvert e_{34}\rvert)=(0.5,0.3)$ with the rest leaf edges and the root edge fixed. Assume $\ba=(-1,-1.2)$ with $\epsilon=1$. We see that the proposed branch lengths are negative as $(\lvert e_{12}\rvert+\epsilon a_1,\lvert e_{34}\rvert+\epsilon a_2)=(-0.5,-0.9)$ with $(\epsilon_1^0,\epsilon_2^0)=(0.5,0.25)$. When $t_\epsilon=0.25$, we obtain $(\lvert e_{12}\rvert,\lvert e_{34}\rvert)=(0.25,0)$ in the codimension of one with three candidate splits of $(e_{34},e_{123},e_{124})$. After excluding the current split of $e_{34}$, assume that $e_{124}$ is chosen. We assign $a_2=1.2$ to the new split of $e_{124}$. When $t_\epsilon=0.5$, we see that $(\lvert e_{12}\rvert,\lvert e_{124}\rvert)=(0.25,0) + 0.25 (-1,1.2) = (0,0.3)$ with the candidate splits of $(e_{24},e_{14})$. Assume $e_{24}$ is selected with $a_1=1$. The final proposed edge set and the branch lengths are $(\lvert e_{24}\rvert,\lvert e_{124}\rvert)=(0.5,0.9)$ at $t_\epsilon=1$.

With the nature of discrete orthants, the derivative of the potential energy is discontinuous and results in a low acceptance rate for HMC \citep{ppHMC_Dinh2017}. One potential remedy to the issue is to approximate the derivative via a smooth surrogate function \citep{Strathmann2015}. For brevity and interest of exposition, we provide a brief description of the surrogate HMC proposed by \citet{ppHMC_Dinh2017}, and refer the reader to the original paper for details. Let surrogate Hamiltonian of $\tilde{H}(\bSigma^T,\ba)=\tilde{U}(\bSigma^T)+K(a)$ with the potential energy as
\begin{align*}
    \tilde{U}(\bSigma^T)=U(\tilde{\bSigma}^T), \thickspace \tilde{\bSigma}^T=\sum_{j=1}^Q g(d_j)\bv_j\bv_j^\transp
\end{align*}
where $g(d)$ is a positive and smooth approximation of the $\lvert d\rvert$ with a gradient vanishing at $d=0$. One common choice of the approximating function is
\begin{align*}
    g_\delta(d)=
    \begin{cases}
        d & \textrm{if } d\geq \delta\\
        \frac{1}{2\delta}(d^2+\delta^2) & \textrm{if } 0\leq d <\delta
    \end{cases},
\end{align*}
where $\delta$ is the smoothing threshold. Obviously, the gradient of $\tilde{U}$ is continuous across orthants due to the vanishing gradient at $d=0$.

\section{Additional Results for Simulation Studies}\label{supp:AddSupp}

\subsection{Convergence diagnostics}
\begin{figure}
    \centering
    \includegraphics[width=\textwidth]{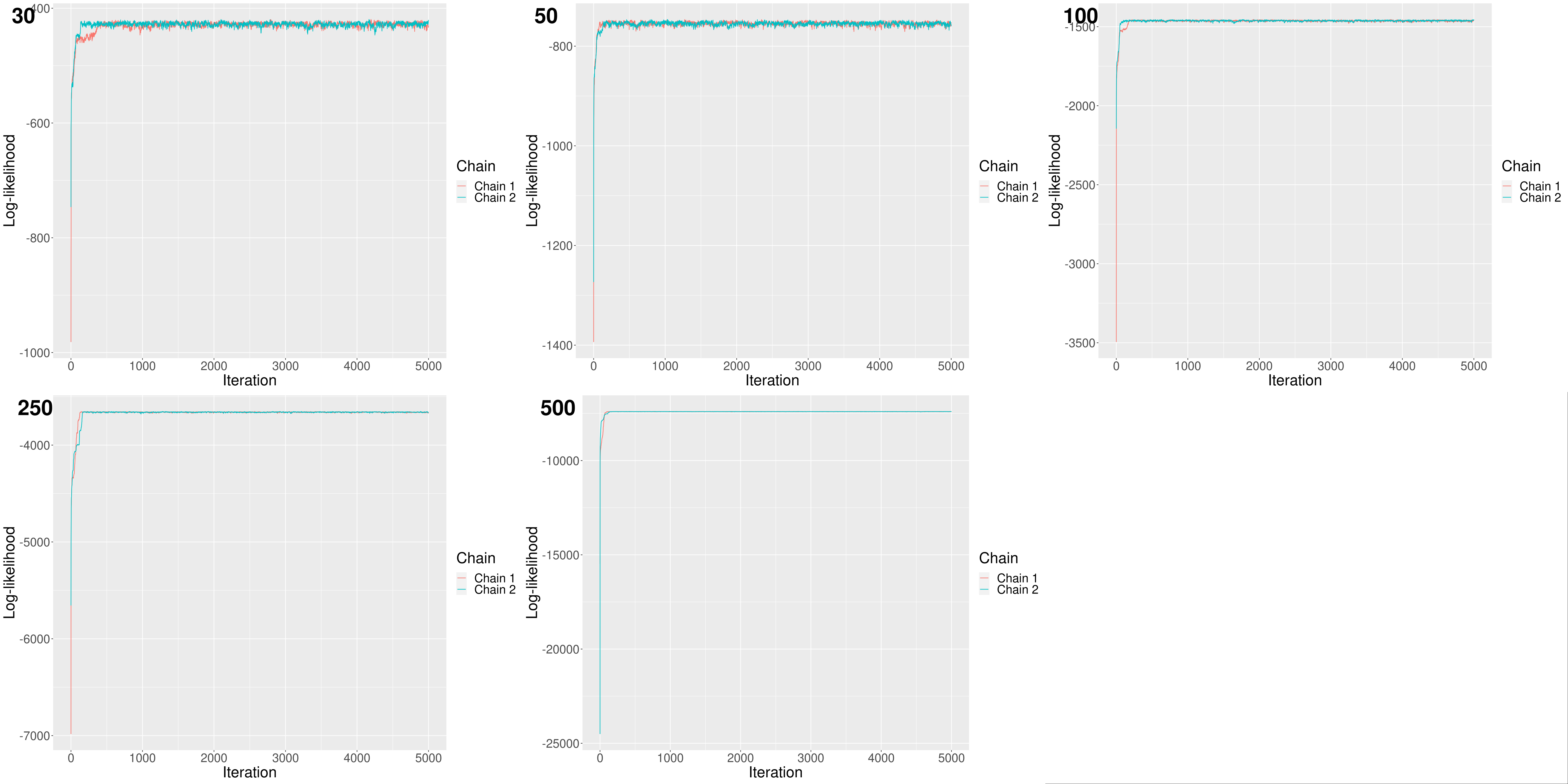}
    \caption{Convergence diagnostics for the algorithm using the likelihood trace plot. Two chains of the same algorithm are initiated by different trees, shown by two colors, across five different sample sizes of $n\in\{30,50,100,250,500\}$.}
    \label{sfig:likelihoodTrace}
\end{figure}

We examine the convergence of our algorithm in Section 6 of Main Paper through the likelihood trace plot and ensure convergence likelihood from different initializations. Specifically, given the same dataset, we run the same algorithm with two chains initiated by two different trees and plot the likelihood over iterations. We randomly chose five likelihood trace plots, each from different sample sizes of $n\in\{30,50,100,250,500\}$, as shown in Figure \ref{sfig:likelihoodTrace}. Obviously, the likelihood increases rapidly and remains at a relatively high plateau. More importantly, both chains converge to a similar level of likelihood, indicating the convergence of the algorithm.

\subsection{Comparing to other proposal functions}\label{supp:DiffAlg}
\begin{figure}[htb!]
    \centering
    \includegraphics[width=\textwidth]{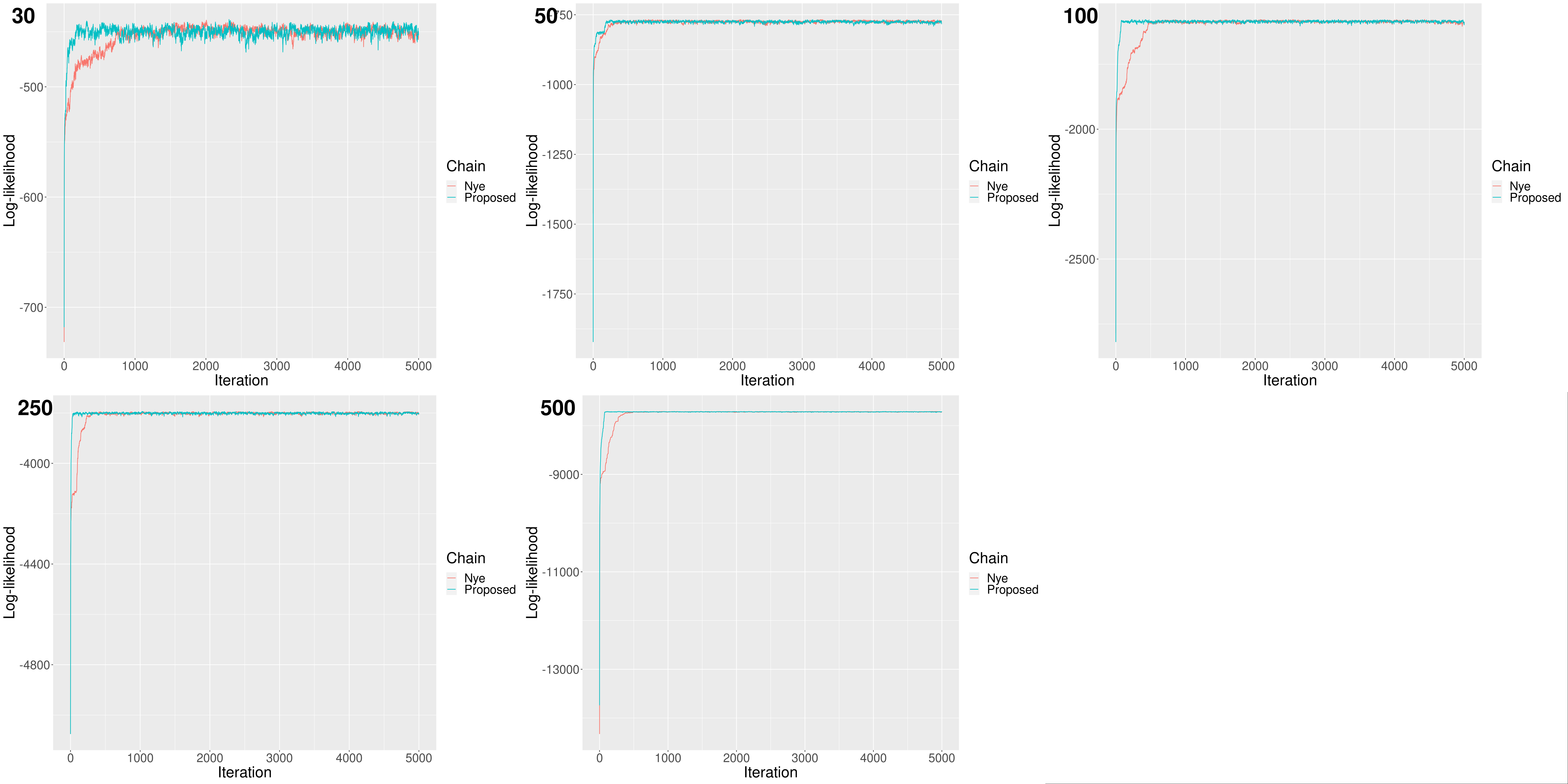}
    \caption{Empirical comparison of our proposed algorithm (blue) and the algorithm from \citet{Nye2020} (red) in terms of the rate of convergence under five different sample sizes of $n \in \{30,50,100,250,500\}$.}
    \label{sfig:cmpAlg}
\end{figure}

We compare our algorithm to the algorithm from \citet{Nye2020} and empirically investigate the rate of convergence through the likelihood trace plot shown in Figure \ref{sfig:cmpAlg}. The algorithm from \citet{Nye2020} enables the algorithm to propose a candidate edge set that is the same as the edge set from the previous iteration, resulting in slower convergence. We briefly describe Nye's algorithm and refer the reader to the original paper for more details. Essentially, Nye's algorithm is still an MH update and changes the proposal function illustrated in Step 1 to 4 in Algorithm 2. Instead of directly removing the internal split in Step 1, Nye's algorithm generates an edge length difference from a normal distribution of $\delta \sim N(0,\sigma_\cE)$ and lets $c = \lvert e_A \rvert + \delta$. When $c > 0$, the algorithm will stay in the same edge set with $e_B=e_A$ and $\lvert e_B \rvert=c$. Otherwise, when $c \leq 0$, the algorithm will propose a candidate split $e_B$ from nearby orthants and assign the edge length of $\lvert e_B \rvert = -c$. However, Nye's algorithm does not exclude the original split from the candidates, resulting in a positive probability of staying in the same edge set and, therefore, slower convergence. We implement Nye's algorithm and compare it to our algorithm. Figure \ref{sfig:cmpAlg} empirically compares the rate of convergence. Obviously, our algorithm converges faster than Nye's algorithm under five different sample sizes.

\subsection{Simulation results from Hamiltonian Monte Carlo}\label{supp:HMC_SimRes}
We empirically compare the HMC and the MH algorithm in terms of the convergence rate and the quality of posterior samples. Following the data-generating mechanism in Section 7, we generate the resolved true tree $T^0$ of $p=10$ leaves and map it to an ultrametric matrix $\bSigma^{T^0}$. We simulate the data on the correctly specified normal distribution $X_i \overset{i.i.d.}{\sim} N(0,\Sigma^{T^0})$ with the sample size $n=25p$ over $30$ replicates. We measure the convergence rate based on the minimum iteration number to achieve a $99\%$ of the maximum of the posterior distribution and the performance of the sampler in the Frobinus norm between the posterior matrices and the $\bSigma^{T^0}$. For HMC algorithm, we let $\delta=0.003$ with the step size $\epsilon=0.0015$ and the number of leap-prog steps of $200$. We ran both algorithm for $300$ iterations and discarded the first $225$ samples.

\begin{figure}
    \centering
    \includegraphics[width=\linewidth]{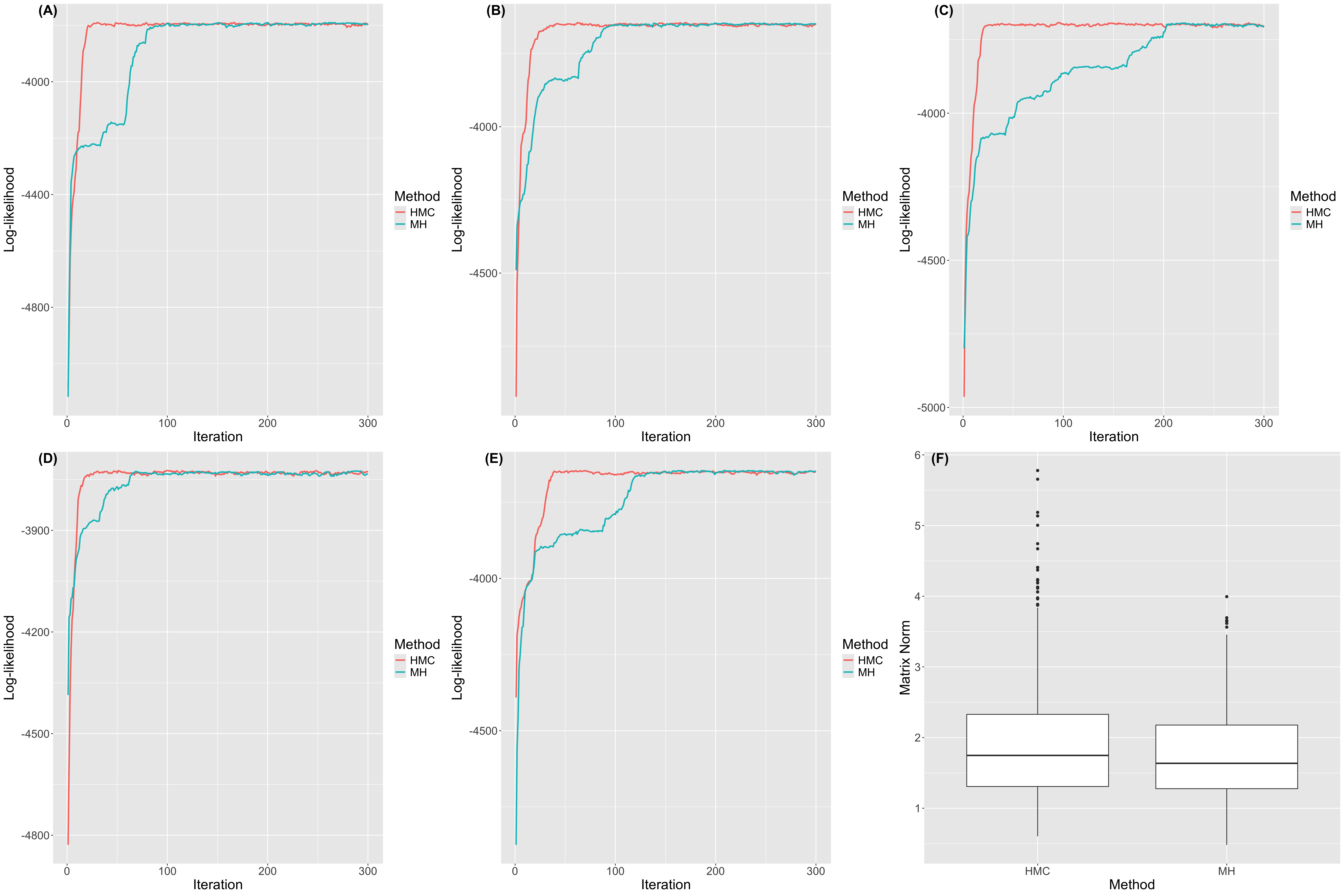}
    \caption{Panel (A) to (E): log-likelihood of the HMC (red) and MH (blue) from five random independent replicates generated from normal distribution based on true a binary tree of $p=10$ leaves. (F): matrix norms between the posterior matrices and the true matrix sampled from HMC and MH algorithms based on $30$ independent replicates.}
    \label{fig:hmc_llh}
\end{figure}

Figure \ref{fig:hmc_llh} shows the trace-plot of the log-likelihood for HMC (red) and MH (blue) algorithms from five randomly selected replicates in Panels (A) to (E) and the quality of posterior matrices in Panel (F). Obviously, HMC algorithm converges much faster ($<35$ iterations) than the MH algorithm ($>63$ iterations). Also, both HMC and MH algorithm achieve the same levels of the log-likelihood, indicating both algorithms converge to around the same level of posterior density. For the performance of the sampler, both algorithms generate a similar quality of posterior matrices in the matrix norm (median of HMC: $1.75$ and MH: $1.64$). In sum, we see that the HMC converges faster than MH algorithm while maintains the quality of the posterior samples.

\subsection{Additional simulation results for resolved tree}\label{supp:AddRes_rtree}
In this Section, we show more results of the uncertainty quantification of the resolved tree. First, the split-wise recovery of the resolved tree is presented in Table \ref{stab:SpltRecov}.

\begin{table}[!htb]
    \caption{Split-wise recovery for the proposed algorithm with $\textrm{exp}(1)$ prior on the edge lengths under a true resolved tree with different sample size, data generating distribution. Each column shows proportion of posterior splits that contains specific split in the true underlying matrix. The average and the standard deviation of the proportion are obtained from $10,000$ iterations with first $9,000$ iterations discarded over $50$ independent replicates.}
    \label{stab:SpltRecov}
    \centering
    \footnotesize
    \begin{tabular}{l|l|llllllll}
    \toprule
    \makecell{Sample \\Size} & \makecell{Generating \\Distribution} &  \makecell{1,2,3,4,5,\\6,7,8,9} & 1,2,4 & 2,4 & 3,5,6,7,8,9 & 3,5,6,8,9 & 3,9 & 5,6 & 5,6,8\\ 
    \midrule
    \multicolumn{2}{c|}{True Edge Lengths} & 0.701 & 0.872 & 0.712 & 0.880 & 0.878 & 0.854 & 0.231 & 0.869 \\
    \midrule
    \multirow[c]{3}{*}{30} & Normal & 79.7(23.7) & 67.8(28.1) & 82.3(20.3) & 80.3(22.6) & 83.2(22.7) & 81.5(20.5) & 43.8(27.6) & 78.7(26.9) \\
    & $t_4$ & 60.4(35) & 72.2(32.6) & 70.8(29.6) & 55.5(37.8) & 65.1(39.4) & 68.4(33.1) & 37(26.9) & 74.2(31.7) \\
    & $t_3$ & 59(38.8) & 60.7(36.2) & 72.2(30.2) & 66.6(38.1) & 68.5(36.7) & 68(33.8) & 42.9(33.7) & 67.1(37.7) \\
    \midrule
    \multirow[c]{3}{*}{50} & Normal & 86.1(18.8) & 90.6(16.6) & 94.8(10.1) & 87.6(19) & 90.5(18.4) & 88.5(22) & 67.8(25.2) & 95(12.8) \\
    & $t_4$ & 78(26.4) & 82.8(27.5) & 78.7(28) & 76.8(30.4) & 76.1(31.8) & 85.9(23.2) & 69.2(30.2) & 90.2(22.9) \\
    & $t_3$ & 72.9(31.5) & 78.3(32.7) & 85.7(23.3) & 72.3(36.9) & 75.4(37.6) & 79.5(32.5) & 46.8(36.4) & 77.6(32) \\
    \midrule
    \multirow{3}{*}{100} & Normal & 95.2(9) & 99.6(1) & 98.6(4.4) & 95.1(11.1) & 98.5(4.3) & 98.7(4.5) & 87.2(17.9) & 99.9(0.3) \\
    & $t_4$ & 88.5(20) & 93.2(18.3) & 96.1(13.2) & 87.5(23.8) & 90(25.3) & 95.2(13.9) & 74.7(34.5) & 93.2(23.9) \\
    & $t_3$ & 71.4(37.3) & 80.4(37.4) & 80.9(37) & 75.9(38.9) & 76.1(39.5) & 87.3(30.4) & 62.8(40.8) & 83.2(35) \\
    \midrule
    \multirow{3}{*}{250} & Normal & 99.6(3) & 100(0.1) & 100(0) & 99.9(0.3) & 100(0) & 100(0) & 98.3(6.4) & 100(0) \\
    & $t_4$ & 95.7(17.4) & 100(0.1) & 99.5(3.1) & 98.9(6.7) & 100(0.1) & 97.9(14.1) & 89.7(20.3) & 100(0) \\
    & $t_3$ & 79.3(38) & 90.4(26.1) & 89.6(29.6) & 90.7(28.5) & 87.8(31.3) & 93.2(23.5) & 86.8(30.3) & 93(24) \\
    \midrule
    \multirow{3}{*}{500} & Normal & 100(0) & 100(0) & 100(0) & 100(0) & 100(0) & 100(0) & 100(0) & 100(0) \\
    & $t_4$ & 100(0) & 98.1(13.4) & 100(0) & 100(0) & 100(0) & 100(0) & 95.8(16) & 100(0) \\
    & $t_3$ & 92.9(24.8) & 91.1(26.3) & 94.1(23.4) & 92.3(26.4) & 95.5(19.8) & 98(14.1) & 88.2(30.6) & 98(14.1) \\
    \bottomrule
    \end{tabular}
\end{table}

\begin{figure}[htb!]
    \centering
    \includegraphics[width=\textwidth]{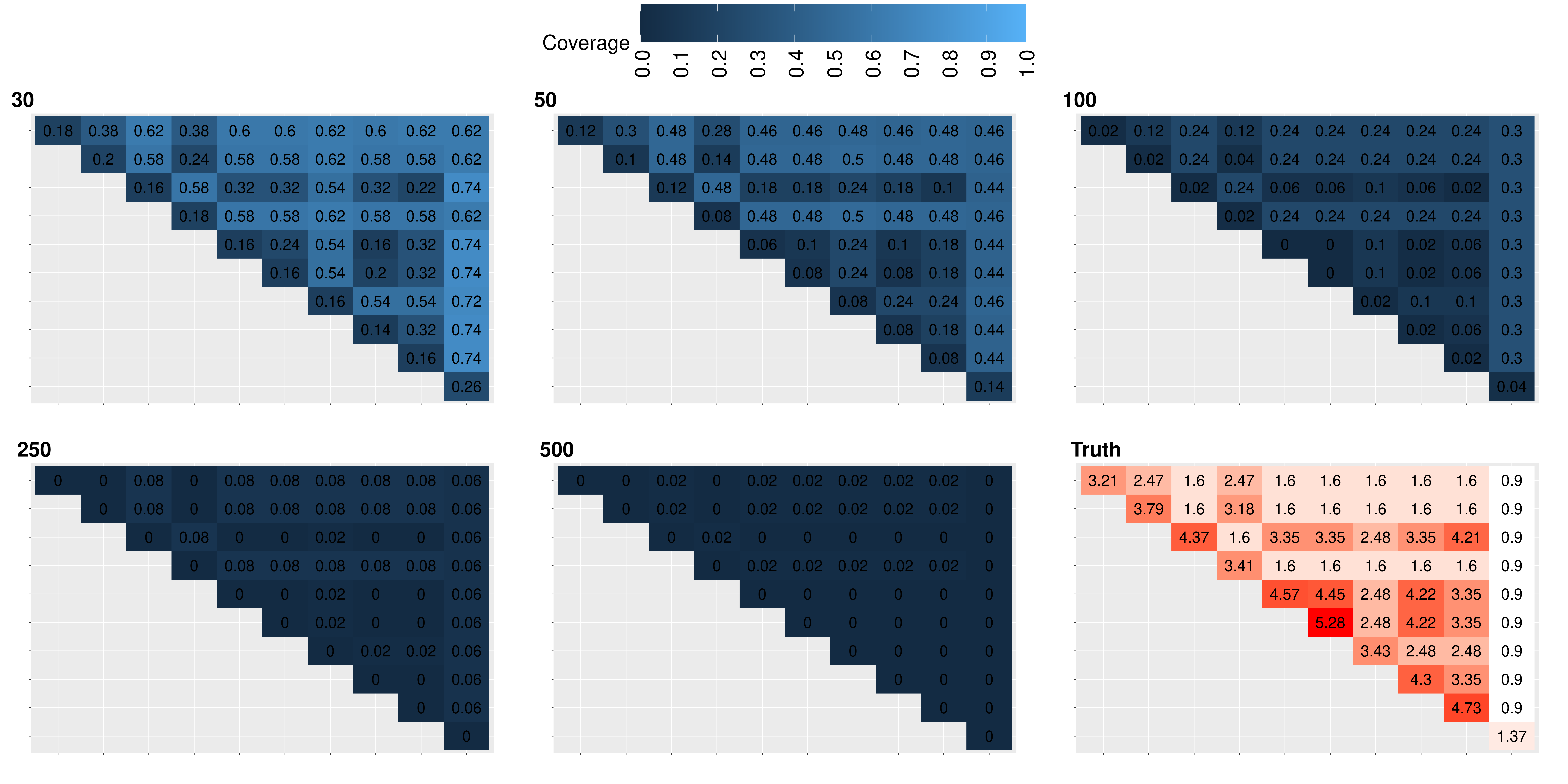}
    \caption{Element-wise coverage from the $95\%$ credit interval for the mis-specified t-distribution of degree of freedom four under five different sample sizes. The true underlying covariance is shown in the lower right panel.}
    \label{sfig:eleCovg_t4}
\end{figure}

\begin{figure}[htb!]
    \centering
    \includegraphics[width=\textwidth]{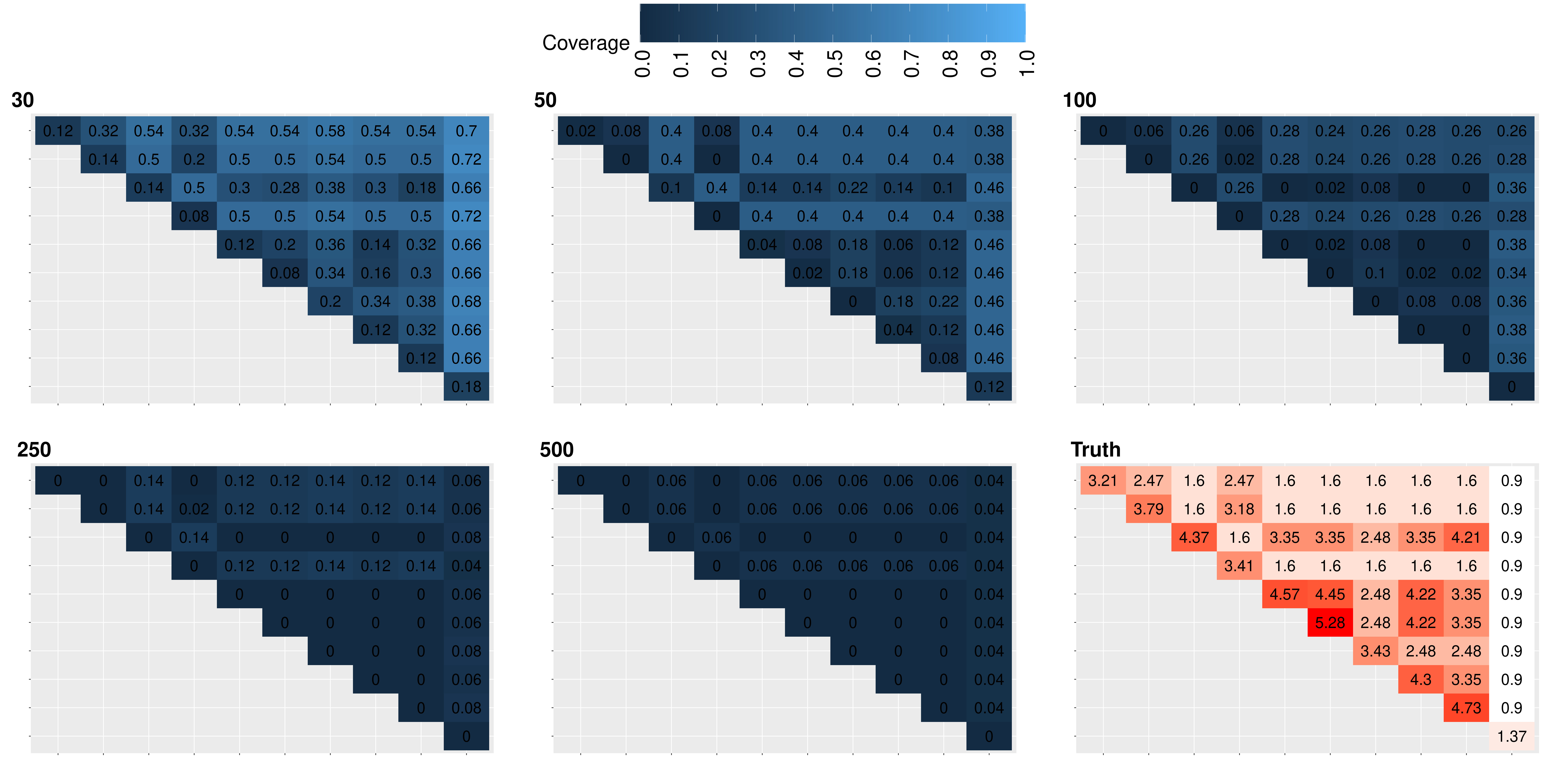}
    \caption{Element-wise coverage from the $95\%$ credit interval for the mis-specified t-distribution of degree of freedom three under five different sample sizes. The true underlying covariance is shown in the lower right panel.}
    \label{sfig:eleCovg_t3}
\end{figure}

We show the nominal coverage for element-wise $95\%$ credible intervals under the mis-specified t-distribution. Specifically, the results of nominal coverage for $t_4$ and $t_3$ for five different sizes are shown in Figure \ref{sfig:eleCovg_t4} and \ref{sfig:eleCovg_t3}, respectively. 
We observe that the nominal coverages for t-distribution are moderate when the sample size is small ($n=30$). However, when the sample size increases, the nominal coverage worsens. Unfortunately, when our algorithm is mis-specified, it does not generate posterior matrices that converge to the true matrix. We suspect that our algorithm converges to incorrect edge lengths when the model is mis-specified. This conjecture is supported by Table 1 in the Main Paper, which indicates that our algorithm still generates posterior samples with the correct topology for t-distribution when the sample size is larger with $n=100$ for $t_4$ and $n=250$ for $t_3$.

\subsection{Topology trajectory for the proposed method}
In this section, we examine the trajectory of our proposed algorithm in $\cU_p$. Specifically, we initiate our algorithm with matrices that are far away (in terms of distance $d$ of Equation (3) in the Main Paper) from the true matrix and track the topologies generated by our algorithm over iterations.

Consider a dataset of sample size $n=500$ that is generated from an underlying normal distribution with a true ultrametric matrix of dimension $p=10$. For the ultrametric matrices of $p=10$, over 34 million possible topology exists, and we index each topology with a natural number. We initiate our algorithm with trees that share no common split with the true matrix and run the algorithm with correct specified normal likelihood for 10,000 iterations. 

Figure \ref{sfig:diffTplg} presents the trajectory of our algorithm applied on the same dataset with $15$ different initial trees in terms of the distance $d$ between the estimated matrices and the true matrix. Each Panel is initiated by the matrix with different topology. For example, Panels in the top row shows the results for the algorithm initiated by matrices with five different topology $2, 27,52, 72$ and $91$, from the left to right. For each Panel, different colors represent different topologies. Obviously, all initial matrices are distant from the true matrix with a higher distance $d$. Over iterations, our algorithm traverses different nearby orthants and quickly moves to the correct topology (topology 1) with a smaller distance $d$ to the true matrix. For example, the top-left Panel shows the distance $d$ for the algorithm initiated by the matrix with topology $2$. Since the initial topology $2$ share no common split to the true matrix, a larger distance $d$ is obtained. The algorithm then quick traverses the topologies $3$ to $26$ and lingers in the correct topology $1$ within the first $150$ iterations with a rapid decrease in distance $d$. After reaching the true topology $1$, we observe that our algorithm sometimes moves to the nearby orthants of topologies $25$ and $26$, but moves back to the correct topology $1$ soon after a short period.

\begin{figure}[htb!]
    \centering
    \includegraphics[width=\textwidth]{Figures/sfig_diffTplg_BHV.png}
    \caption{Trajectory of our algorithm in terms of distance $d$. Over iterations, distances $d$ between each posterior matrix and the true matrix are measured. Each posterior sample is colored according to the corresponding topology. The same algorithm is initiated by $15$ different matrices that are far away (in terms of the distance $d$) from the true matrix. Our algorithm traverses different orthants and arrives at the true topology (topology 1) quickly after a few iterations.}
    \label{sfig:diffTplg}
\end{figure}

\subsection{Simulation results for the tree with the same edge lengths from the root}
In this Section, we demonstrate additional simulation results when all leaves in the true underlying tree are equidistant to the root. Specifically, we obtain a tree from the coalescence model implemented by the function \texttt{rcoal} in the \texttt{R} package \texttt{ape} and generate the data from the normal and t-distribution described in the Main Paper Section 6. We run Algorithm 1 without restricting the prior on the edge lengths for $10,000$ iterations and discard the first $9,000$ iterations. We summarize the posterior samples by the point estimation and the quantify the uncertainty through the element-wise $95\%$ credible interval. The performance of the point estimator is compared to the projection-based method of \citet{MIP_Tr} and the sample covariance under the measurement of the distance $d$ from Equation (3) in the Main Paper and the matrix norm. For the uncertainty quantification, we calculate the nominal coverage of the $95\%$ credible interval. All results were obtained from $50$ independent replicates.

\begin{figure}[htb!]
    \centering
    \includegraphics[width=\textwidth]{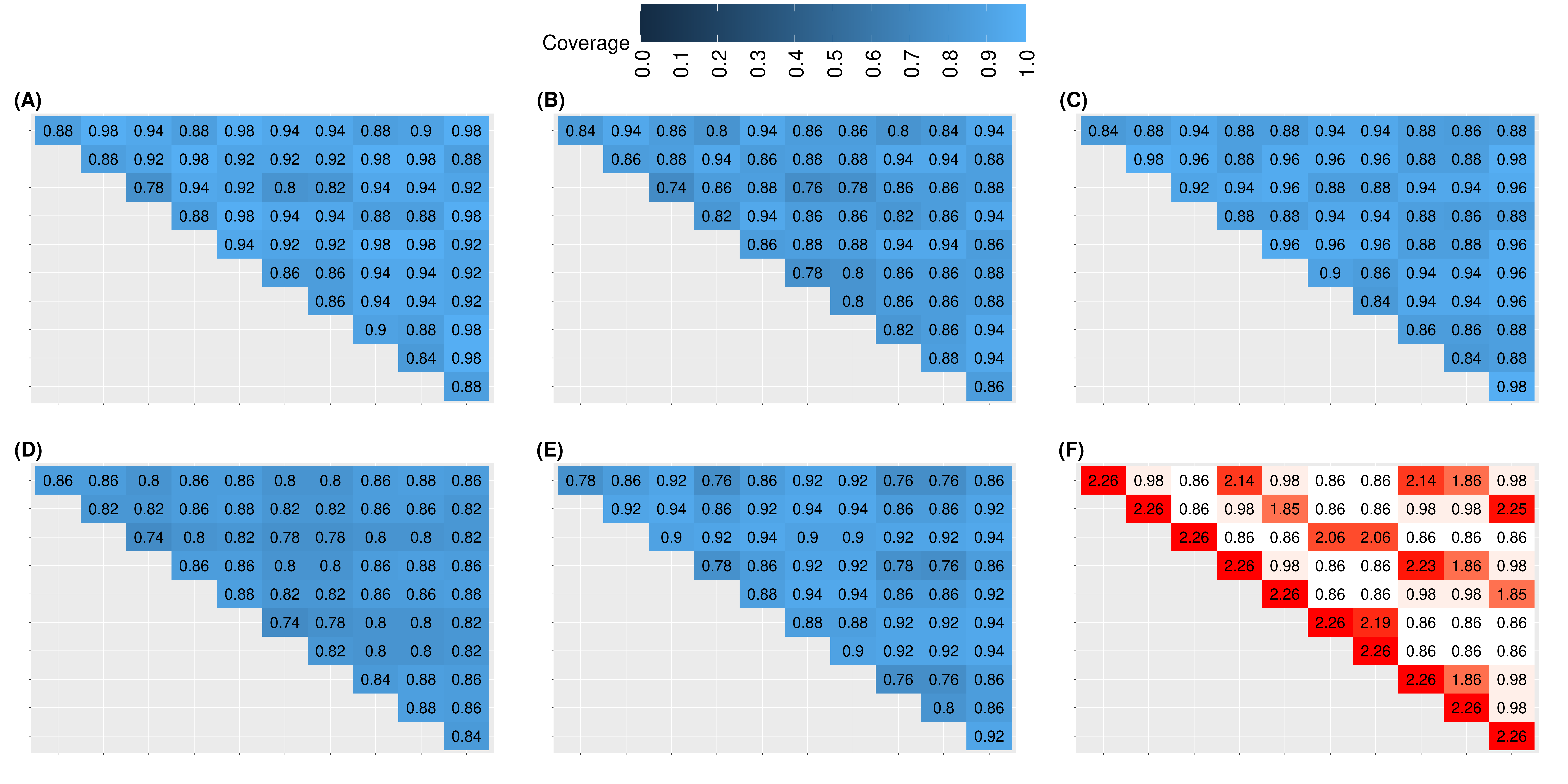}
    \caption{Element-wise coverage from the $95\%$ credit interval for the correct specified normal distribution with five different sample sizes with the true underlying matrix in the Panel (F). Equal diagonal elements in the true matrix indicate that all leaves in the true underlying tree are equidistant to the root.}
    \label{sfig:rcoal_covgEleWise}
\end{figure}

Figure \ref{sfig:rcoal_covgEleWise} demonstrates the nominal coverage of the element-wise $95\%$ credible interval with the true generating covariance in the Panel (F). As we expected, the diagonal elements in the true underlying covariance are equivalent, implying that all leaves in the true underlying tree are equidistant to the root. Similar to the results shown in Main Paper Section 6, the $95\%$ credible interval gives a high nominal coverage (around $0.73$ to $1$), and the estimated coverage is higher when the sample size increases. In summary, our algorithm can efficiently draw posterior samples of matrices under different conditions imposed on the edge lengths of the true underlying tree.

The results of the point estimators are shown in Figure \ref{sfig:rcoal_PtDist}. For the point estimator, the mean and MAP trees from our method are comparable to the estimated matrix from \citet{MIP_Tr} and sample covariance in terms of distance $d$ and matrix norm across different data generating mechanisms and sample sizes. When the model is correctly specified, all methods benefit from the increase in the sample size, resulting in a smaller distance to the true matrix. For the mis-specified scenario, the advantage from the larger sample size is moderate. Essentially, when all leaves in the true underlying tree are equidistant to the root, our algorithm still generates posterior samples that are comparable to existing methods with a similar level of performance in terms of the distance $d$ and matrix norm.

\begin{figure}[htb!]
    \centering
    \includegraphics[width=\textwidth]{Figures/sfig_rcoal_nLeaf10_ResPlot.png}
    \caption{Distances between the estimated matrix and the true matrix under different data generating mechanism and sample sizes. The mean (red) and MAP tree (green) from our method is comparable to competing methods (blue for MIP and purple for sample covariance) in terms of the distance $d$ (top row) and matrix norm (bottom row).}
    \label{sfig:rcoal_PtDist}
\end{figure}

\subsection{Scalability of the proposed method}\label{supp:HigherP}
We assess the scalability of the proposed method by considering a larger dimension of $p\in\{10,20,30\}$ for the true matrix $\Sigma^{T^0}$. We repeat the same data generating mechanism of the normal distribution of $X_i \overset{i.i.d.}{\sim} N(0,\Sigma^{T^0})$ with $i=1,\ldots,n$ and $n=50p$. We measure the running time of the proposed Algorithm with $10,000$ iterations and compare it to the running time of the competing method MIP \citep{MIP_Tr}. All results are executed on Linux-based cluster with a CPU 2x 3.0 GHz Intel Xeon Gold 6154 and varying memory sizes required for different methods. Specifically, we assign 1GB of memory for the proposed method and 25GB of memory to MIP.

\begin{figure}
    \centering
    \includegraphics[width=0.5\textwidth]{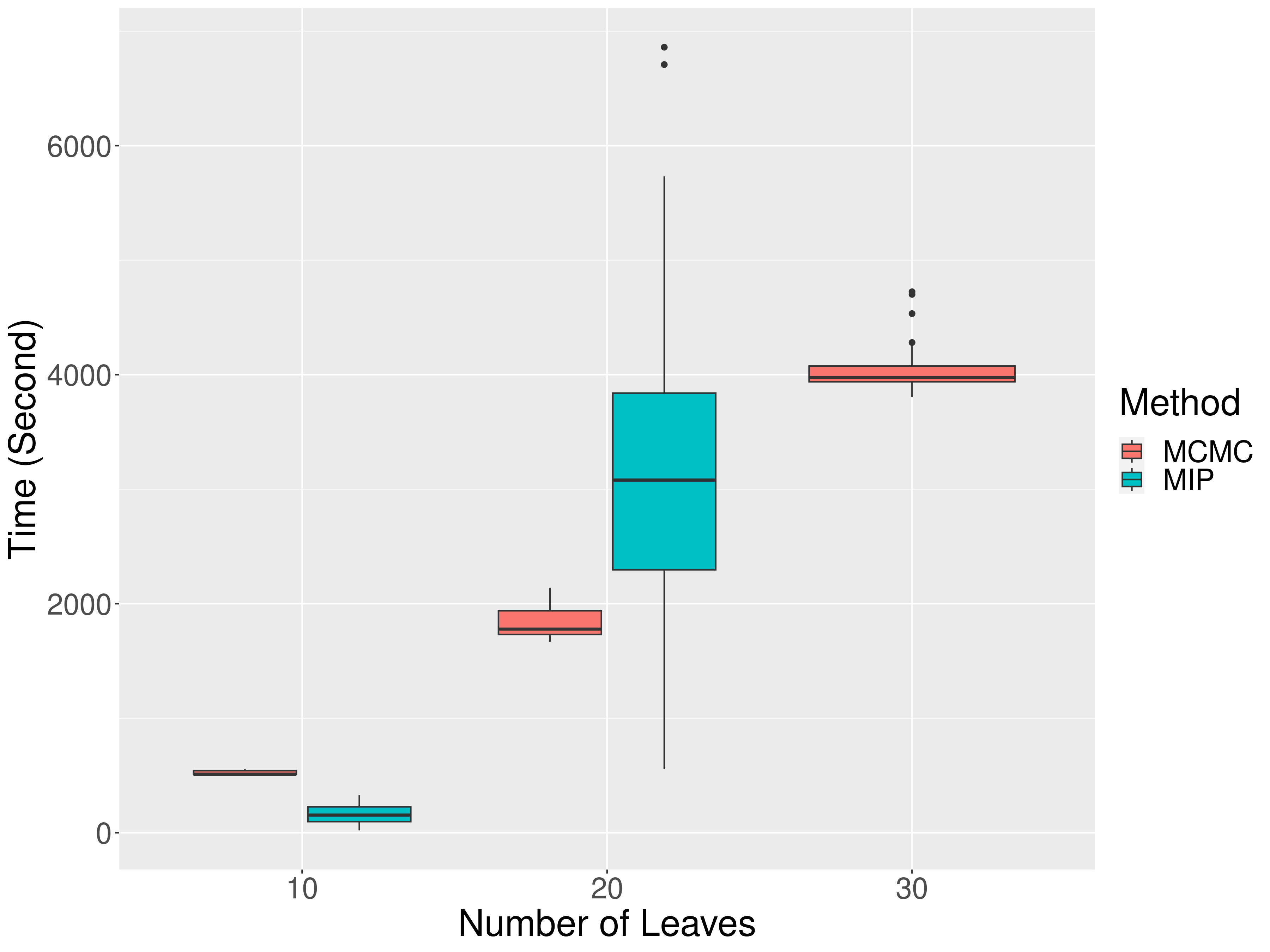}
    \caption{Computation time (in seconds) required for the proposed method (MCMC) and the competing method (MIP) under three different number of leaves $p\in \{10, 20, 30\}$. The computation time of MIP for $p=30$ is missing for the MIP fails after running for $3$ days due to the memory and computation limits.}
    \label{sfig:runTime}
\end{figure}

The results of the running time are shown in Figure \ref{sfig:runTime}. For $p=10$, MIP is faster than the proposed MCMC (median running time in seconds for MIP: $154$ and MCMC: $511$). However, when considering a larger dimension of $p=20$, the proposed MCMC takes around $30$ minutes (median: $1,788$ seconds), which is $1.75$ times faster than the time required for MIP (median: $3,080$ seconds). If we increase the dimension to $p=30$, the proposed method spends $66$ minutes (median: $3,976$ seconds), while MIP fails after running for $3$ days due to memory and computation limits. In conclusion, our method scales better in $p$ with a much efficient memory requirement.

\section{Additional Results for the Real Data Application}\label{supp:RealDataConvg}
We run the convergence diagnostics of our algorithm through the likelihood trace plot. Specifically, given the PDX dataset, we run the same algorithm with two chains initiated by two different trees and plot the likelihood over iterations. When the algorithm is converged, the likelihood initiated by two different chain should reach 
a similar level. Obviously, the likelihood increases rapidly and remains at a relatively high plateau. More importantly, both chains converge to a similar level of likelihood, indicating the convergence of the algorithm.

\begin{figure}
    \centering
    \includegraphics[width=0.5\textwidth]{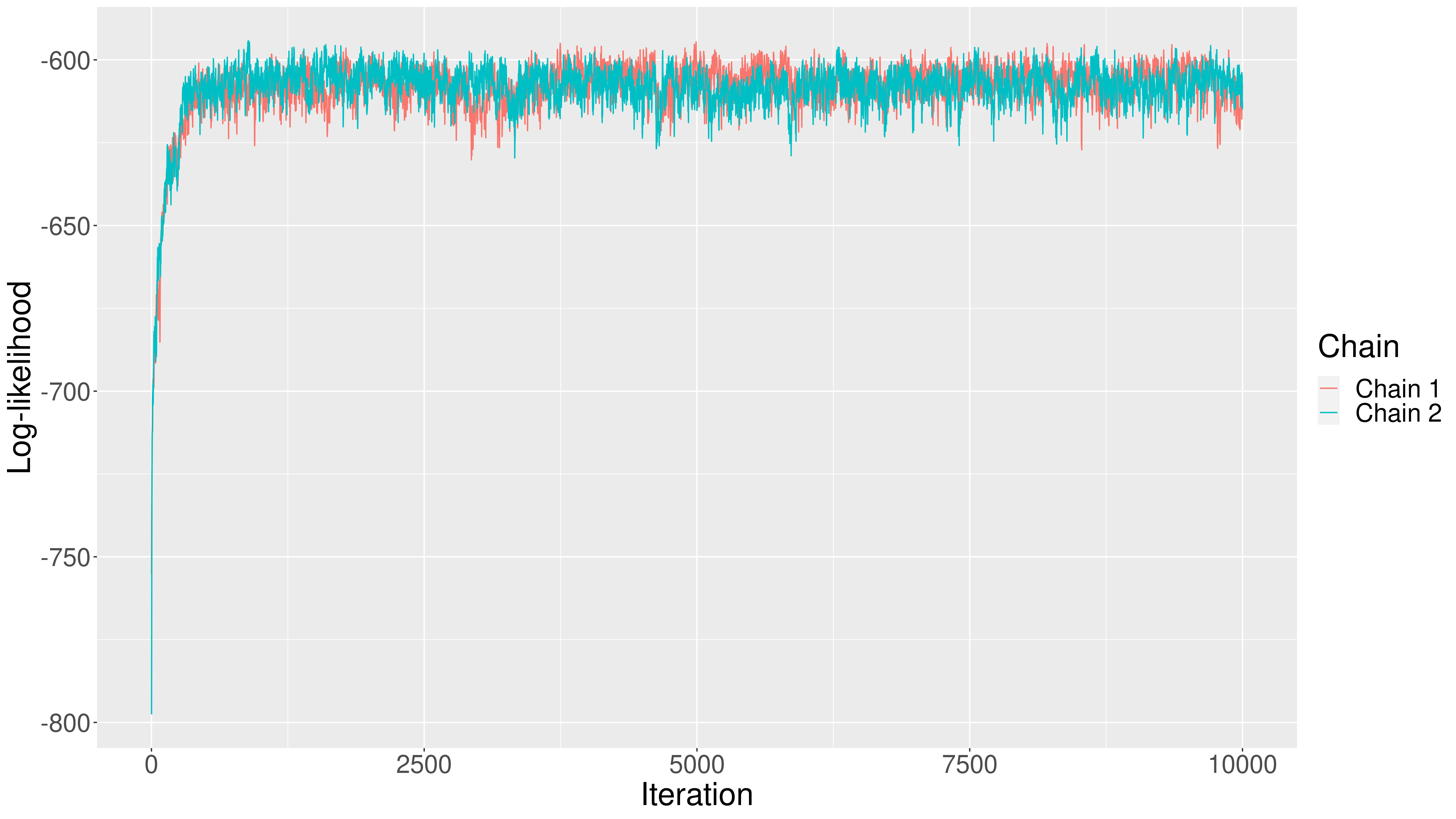}
    \caption{Convergence diagnostics for the algorithm on PDX dataset by using the likelihood trace plot. Two chains of the same algorithm are initiated by different trees, shown by two colors.}
    \label{sfig:realDataConv}
\end{figure}

\clearpage
\FloatBarrier
\bibliography{UltraMat}

%% file: definitions.tex
\def\0{\mbox{\boldmath{$\mathbf{0}$}}}
\def\1{\mbox{\boldmath{$\mathbf{1}$}}}
\def\bzeta{\mbox{\boldmath$\zeta$}}
\def\bnu{\mbox{\boldmath$\nu$}}
\def\btheta{\mbox{\boldmath$\theta$}}
\def\bTheta{\mbox{\boldmath$\Theta$}}
\def\bmu{\mbox{\boldmath$\mu$}}
\def\bbeta{\mbox{\boldmath$\beta$}}
\def\bchi{\mbox{\boldmath$\chi$}}
\def\boldeta{\mbox{\boldmath$\eta$}}
\def\bzeta{\mbox{\boldmath$\zeta$}}
\def\bepsilon{\mbox{\boldmath$\epsilon$}}
\def\bomega{\mbox{\boldmath$\omega$}}
\def\bOmega{\mbox{\boldmath$\Omega$}}
\def\bgamma{\mbox{\boldmath$\gamma$}}
\def\bGamma{\mbox{\boldmath$\Gamma$}}
\def\bsigma{\mbox{\boldmath$\sigma$}}
\def\bSigma{\mbox{\boldmath$\Sigma$}}
\def\bdelta{\mbox{\boldmath$\delta$}}
\def\bDelta{\mbox{\boldmath$\Delta$}}
\def\blambda{\mbox{\boldmath$\lambda$}}
\def\bLambda{\mbox{\boldmath$\Lambda$}}
\def\calF{\mbox{$\mathcal{F}$}}
\def\calH{\mbox{$\mathcal{H}$}}
\def\calI{\mbox{$\mathcal{I}$}}
\def\calJ{\mbox{$\mathcal{J}$}}
\def\calX{\mbox{$\mathcal{X}$}}
\def\A{\mbox{\boldmath{$\mathbf{A}$}}}
\def\B{\mbox{\boldmath{$\mathbf{B}$}}}
\def\C{\mbox{\boldmath{$\mathbf{C}$}}}
\def\D{\mbox{\boldmath{$\mathbf{D}$}}}
\def\G{\mbox{\boldmath{$\mathbf{G}$}}}
\def\I{\mbox{\boldmath{$\mathbf{I}$}}}
\def\J{\mbox{\boldmath{$\mathbf{J}$}}}
\def\M{\mbox{\boldmath{$\mathbf{M}$}}}
\def\N{\mbox{\boldmath{$\mathbf{N}$}}}
\def\R{\mbox{\boldmath{$\mathbf{R}$}}}
\def\S{\mbox{\boldmath{$\mathbf{S}$}}}
\def\U{\mbox{\boldmath{$\mathbf{U}$}}}
\def\X{\mbox{\boldmath{$\mathbf{X}$}}}
\def\W{\mbox{\boldmath{$\mathbf{W}$}}}
\def\Y{\mbox{\boldmath{$\mathbf{Y}$}}}
\def\Z{\mbox{\boldmath{$\mathbf{Z}$}}}
\def\a{\mbox{\boldmath{$\mathbf{a}$}}}
\def\c{\mbox{\boldmath{$\mathbf{c}$}}}
\def\e{\mbox{\boldmath{$\mathbf{e}$}}}
\def\f{\mbox{\boldmath{$\mathbf{f}$}}}
\def\h{\mbox{\boldmath{$\mathbf{h}$}}}
\def\j{\mbox{\boldmath{$\mathbf{j}$}}}
\def\p{\mbox{\boldmath{$\mathbf{p}$}}}
\def\x{\mbox{\boldmath{$\mathbf{x}$}}}
\def\y{\mbox{\boldmath{$\mathbf{y}$}}}
\def\z{\mbox{\boldmath{$\mathbf{z}$}}}